# OpenGLSA: A high-order matrix-free solver for global linear stability analysis of compressible flows on structured grids

Rui-Qing Yang, Bao-Feng Ma*

*Key Laboratory of Fluid Mechanics (Ministry of Education), School of Aeronautic Science and Engineering, Beihang University, Beijing 102206, China*

## Abstract

We present OpenGLSA, an open-source code for the global linear stability analysis of compressible flows. It is based on a high-order finite-difference method on structured curvilinear grids and parallelised via the Message Passing Interface (MPI). The code supports linear stability analysis around planar and axisymmetric two-dimensional base flows with three-dimensional perturbations represented by Fourier mode expansions (BiGlobal), and on fully three-dimensional base flows (TriGlobal). Curvilinear coordinate grids are employed to facilitate flow analysis around complex geometries. Base flow computations employ the Selective Frequency Damping (SFD) method combined with weighted essentially non-oscillatory (WENO) reconstruction for shock capturing. A linearise-then-discretise approach is adopted for the linearised equations, and eigenmodes are computed using matrix-free time stepping coupled with the implicitly restarted Arnoldi method. This strategy significantly reduces memory requirements, thereby enhancing both the efficiency and tractability of memory-intensive compressible flow analyses. The solvers are validated against published results for low-Mach-number flows past a circular cylinder and a sphere, hypersonic flow over a compression ramp, and square and cubic lid-driven cavity flows.

## Program summary

*Program title*: **OpenGLSA**

*Developer's repository link:* https://github.com/Dxslab/OpenGLSA.git

*Licensing provisions:* Apache License 2.0

*Programming language:* Fortran

* Corresponding author.

*E-mail address*: bf-ma@buaa.edu.cn

*Nature of problem:*
OpenGLSA solves the nonlinear and linearised compressible Navier–Stokes equations in two and three dimensions, and computes global linear eigenmodes via Biglobal and Triglobal stability analyses for both smooth and shock-dominated flows.
*Solution method:*
The governing equations are discretized using high-order finite-difference schemes on structured grids. Specifically, a fifth-order WENO nonlinear reconstruction is employed for the nonlinear equations, while a fifth-order linear upwind reconstruction is used for the linearised equations. Viscous terms are approximated by central differences, and time integration is performed via a third-order Runge–Kutta scheme. For the linearised equations, a linearise-then-discretise approach is adopted; global linear eigenmodes are subsequently computed using matrix-free time stepping coupled with the implicitly restarted Arnoldi method. Flow analysis around complex geometries is conducted using curvilinear coordinate grids.
*Additional comments including restrictions and unusual features:*
The code uses single-block structured curvilinear meshes. The stability analysis is restricted to steady base flows. MPI and ARPACK/PARPACK are required.

# 1. Introduction

Complex flow phenomena in many engineering applications originate from flow instabilities. Stability analysis serves as a vital tool for revealing the underlying mechanisms of these complex flows, thereby facilitating the development of effective flow control strategies. Global linear stability analysis (GLSA) provides a framework for investigating the growth and spatial structure of disturbances at the onset of instability in spatially developing flows [1]. By accounting for the spatial inhomogeneity of the base flow in two or three directions, it extends classical stability analysis beyond the parallel and weakly non-parallel flow assumptions [2]. For a steady base flow, the resulting eigenvalue problem determines the growth rates, frequencies and spatial structures of the global modes, allowing the identification of both stationary and oscillatory instabilities [2,3].

In recent years, GLSA has been widely applied to investigate bifurcations and predict the onset of unsteadiness in compressible flows, particularly in shock-dominated flows. Hildebrand et al. [4] investigated the global linear stability of oblique shock wave/boundary layer interactions at Mach 5.92. Hao et al. [5] examined the onset of three-dimensional instability in hypersonic compression corner flows, while Cao et al. [6] combined GLSA and direct numerical simulations (DNS) to relate the unsteadiness of hypersonic compression-ramp flows to intrinsic global instabilities. For axisymmetric configurations, Li and Hao [7] examined supersonic flow over a hollow cylinder/flare and showed how axisymmetric effects modify flow separation and the onset of global instability. Beyond laminar flows, GSA of Reynolds-averaged Navier-Stokes equations has been applied to investigate transonic aerofoil buffet [8].

These applications highlight the requirement for reliable and computationally accessible tools for compressible GLSA. Such calculations require sufficiently converged base flows and accurate resolution of small-amplitude disturbances, particularly in flows containing shocks and separated regions. The resulting large-scale eigenvalue problems place substantial demands on computational resources, making the choice of solution strategy important.

The numerical methods for GLSA differ in how the perturbation operator is constructed and how its eigenmodes are computed. Linearising the governing equations before discretisation allows distinct numerical schemes to be employed for the nonlinear and perturbation equations; however, this approach necessitates separate implementation and verification of the perturbation solver. Alternatively, differentiating the discrete nonlinear residual yields its Jacobian, thereby preserving consistency with the numerical treatments embedded in the discretisation. Automatic differentiation can generate these derivatives exactly. Finite-difference approximations offer another option by estimating the action of the Jacobian without explicit matrix formation. While this enables reuse of existing nonlinear solvers, it introduces truncation errors that depend on both the perturbation amplitude and the specific difference formula employed [9].

The choice of eigenvalue solver affects both memory usage and computational cost. Matrix-based approaches with shift-and-invert techniques can target eigenvalues near a prescribed location in the spectrum, but storing the matrix and solving the associated linear systems can become expensive as the problem grows [2,9]. Time-stepping approaches evaluate the action of a perturbation propagator through repeated integrations and use Krylov methods to extract its dominant modes [2,10]. They avoid assembling the global Jacobian and are suitable for finding the leading modes that govern the onset of instability. They still require repeated time integrations and storage of the Krylov basis, and their convergence depends on the spectral separation of the modes of interest [9]. Lower storage requirements therefore do not necessarily imply lower computation cost.

Existing open-source codes employ different combinations of these preceding methods. Nektar++ [11], Semtex [12] and nekStab [10,13] provide high-order tools for stability analysis on incompressible flow. For compressible flows, BROADCAST [14] employs algorithmic differentiation for stability, resolvent and sensitivity analyses on two-dimensional base flows, with three-dimensional disturbances represented by spanwise Fourier expansions. Its shift-and-invert algorithm stores the Jacobian and sparse LU factors, making memory a major constraint when extending this approach to fully three-dimensional base flows. HYMOR [15] also assembles a sparse linearised operator, but uses a GPU-accelerated exponential spectral transformation with Arnoldi iteration to avoid shift-and-invert factorisation. Its shock-fitting formulation for the bow shock requires additional treatment of shock surfaces and their intersections to handle complex interacting shock systems.

This paper presents OpenGLSA, an open-source high-order finite-difference solver for the global stability analysis of steady compressible flows on structured curvilinear grids. The governing equations are linearised before discretisation, allowing fifth-order multi-resolution WENO reconstruction [16] for the nonlinear equations and fifth-order linear upwind reconstruction for the linearised equations, equipping the solver to effectively capture shock-dominated flows. It combines a matrix-free time stepping method with the implicitly restarted Arnoldi method. The implicitly restarted Arnoldi method [17] extracts the leading modes from the perturbation propagator without storing the global Jacobian or its factors. The program is parallelised by domain decomposition using the Message Passing Interface (MPI). It supports planar and axisymmetric base flows through Fourier representations of disturbances, reducing these calculations to two-dimensional grids, and includes a fully three-dimensional solver for base flows without these symmetries.

The validation cases comprise cylinder and sphere wakes, hypersonic compression-ramp flow, and two- and three-dimensional lid-driven cavity (LDC) flows. They provide tests of the planar, axisymmetric and fully three-dimensional formulations in smooth and shock-containing configurations. A cylinder-flow comparison with BROADCAST assesses memory and runtime requirements under the reported computational settings.

The rest of the paper is organised as follows. Sec. 2 introduces the governing equations and linear perturbation formulations for planar, axisymmetric and fully three-dimensional flows. Sec. 3 details the spatial discretisation and time-integration schemes. Sec. 4 describes the procedures for computing steady base flows and global modes, together with the software organisation and parallel implementation. Sec. 5 presents the validation cases. The conclusions and future developments are included in Sec. 6.

## 2. Governing equations and perturbation formulation

This section presents the nondimensional Navier-Stokes equations governing fluid flows and corresponding linearised equations. For two-dimensional planar and axisymmetric configurations, Fourier mode decomposition is employed to describe expansions in the spanwise or azimuthal direction.

Cartesian coordinates ($x$, $y$, $z$) are used for planar two-dimensional and three-dimensional configurations. For axisymmetric two-dimensional configurations, cylindrical coordinates ($x$, $y$, $\theta$) are used, where $x$ and $y$ denote the axial and radial coordinates, respectively. The velocity components ($u$, $v$, $w$) correspond to the coordinate directions in each system. Curvilinear coordinate transformations are employed to facilitate flow analysis around complex geometries.

Length, velocity, density and temperature are nondimensionalised using reference length $L_{\text{ref}}$, freestream velocity $U_\infty$, freestream density $\rho_\infty$ and freestream temperature $T_\infty$, respectively, while pressure is nondimensionalised using $\rho_\infty U_\infty^2$. The Reynolds number and the Mach number are defined as $Re_\infty = \rho_\infty U_\infty L_{\text{ref}} / \mu_\infty$ and $Ma_\infty = U_\infty / c_\infty$, where $c_\infty$ is the freestream speed of sound.

### *2.1. Compressible Navier-Stokes equations*

The nondimensional compressible Navier-Stokes equations in Cartesian coordinates are written in conservative form,

$$
\begin{aligned}
&\frac{\partial \rho}{\partial t} + \frac{\partial}{\partial x_k}(\rho u_k) = 0, \\
&\frac{\partial}{\partial t}(\rho u_i) + \frac{\partial}{\partial x_k}(\rho u_i u_k + p\delta_{ik}) = \frac{\partial \tau_{ik}}{\partial x_k}, \\
&\frac{\partial}{\partial t}(\rho E) + \frac{\partial}{\partial x_k}(\rho u_k (E + \frac{p}{\rho})) = \frac{\partial}{\partial x_k}(u_i \tau_{ik} - q_k),
\end{aligned}
\tag{1}
$$

where $q_k$ and $\tau_{ik}$ denote the heat flux and stress tensor, respectively, given by

$$
q_k = -\kappa \frac{\partial T}{\partial x_k}, \tag{2}
$$

$$\tau_{ik} = \mu(\frac{\partial u_i}{\partial x_k} + \frac{\partial u_k}{\partial x_i} - \frac{2}{3}\frac{\partial u_j}{\partial x_j}\delta_{ik}). \quad (3)$$

Here, $\rho$, $u_i$, $p$, $T$ and $E$ are the density, velocity, pressure, temperature, and total specific energy, respectively. The dynamic viscosity $\mu$ is evaluated using the nondimensional form of Sutherland's law,

$$\mu = \frac{1}{Re_\infty}(T)^{3/2}\frac{1+S_\mu}{T+S_\mu}. \quad (4)$$

Here, $S_\mu = 110.4/T_\infty$ is the nondimensional Sutherland constant, with $T_\infty$ denoting the freestream temperature. The thermal conductivity is calculated as

$$\kappa = \frac{\mu}{Pr(\gamma-1)Ma_\infty^2}, \quad (5)$$

where $\gamma$ is the ratio of specific heats and $Pr$ is the Prandtl number. Unless otherwise stated, $\gamma$=1.4 and $Pr$ =0.72 are used for the validation cases presented in this work.

The system is closed by assuming a calorically perfect gas, for which

$$p = \frac{1}{\gamma Ma_\infty^2}\rho T, \qquad E = \frac{p}{\rho(\gamma-1)} + 0.5u_i u_i. \quad (6)$$

In cylindrical coordinates, the viscous stress tensor is expressed as

$$\tau_{ik} = \mu(G_{ik} + G_{ki} - \frac{2}{3}G_{jj}\delta_{ik}), \quad (7)$$

where $G_{ij}$ is the velocity gradient tensor,

$$G_{ij} = \begin{bmatrix} \frac{\partial u}{\partial x} & \frac{\partial u}{\partial y} & \frac{\partial u}{y\partial\theta} \\ \frac{\partial v}{\partial x} & \frac{\partial v}{\partial y} & \frac{\partial v}{y\partial\theta} - \frac{w}{y} \\ \frac{\partial w}{\partial x} & \frac{\partial w}{\partial y} & \frac{\partial w}{y\partial\theta} + \frac{v}{y} \end{bmatrix}. \quad (8)$$

With this stress tensor, the governing equations in cylindrical coordinates retain the form of Eq. (1) under the derivative replacement

$$(\frac{\partial}{\partial x},\frac{\partial}{\partial y},\frac{\partial}{\partial z})\rightarrow(\frac{\partial}{\partial x},\frac{\partial}{\partial y},\frac{\partial}{y\partial\theta}), \tag{9}$$

which also applies to the definition of heat flux. Furthermore, an additional source term for the geometric contributions should be added to the right-hand side of Eq. (1) in cylindrical coordinates:

$$\mathbf{S}_{\text{axis}}=\frac{1}{y}\begin{bmatrix} -\rho v \\ \tau_{12}-\rho uv \\ \tau_{22}-\tau_{33}-\rho v^2+\rho w^2 \\ 2\tau_{23}-2\rho vw \\ u\tau_{12}+v\tau_{22}+w\tau_{23}-q_2-\rho v(E+p/\rho) \end{bmatrix}. \tag{10}$$

## *2.2. Linearised perturbation equations*

The primitive state vector $\mathbf{q}=(\rho, u, v, w, T)^{\mathrm{T}}$ is decomposed into a steady base flow and a small perturbation,

$$\mathbf{q}(x,y,z,t)=\bar{\mathbf{q}}(x,y,z)+\mathbf{q}'(x,y,z,t). \tag{11}$$

where $\mathbf{q}'=(\rho', u', v', w', T')^{\mathrm{T}}$. Substituting Eq. (11) into the governing equations and neglecting higher-order terms yields the linearised perturbation equations. After spatial discretisation, these equations are written as

$$\frac{\mathrm{d}\mathbf{q}'}{\mathrm{d}t}=\mathbf{A}\mathbf{q}', \tag{12}$$

where $\mathbf{A}$ is the discrete linearised operator, including the linearised boundary conditions. Since the governing equations are linearised before spatial discretisation, $\mathbf{A}$ is, in general, not the exact Jacobian of the discrete nonlinear residual.

For a steady base flow, the perturbation admits normal-mode solutions of the form

$$\mathbf{q}'(x,y,z,t)=\mathbf{q}(x,y,z)\exp(\lambda t),\qquad \lambda=\lambda_r+i\lambda_i, \tag{13}$$

where $\hat{\mathbf{q}}$ is the spatial eigenfunction, $\lambda_r$ is the growth rate and $\lambda_i$ is the angular frequency. A mode is unstable when $\lambda_r > 0$ and oscillatory when $\lambda_i \neq 0$.

Substituting Eq. (13) into Eq. (12) yields the eigenvalue problem

$$\mathbf{A}\mathbf{q} = \lambda \mathbf{q} \,. \tag{14}$$

The operator **A** is not assembled explicitly. The action of its associated evolution operator is evaluated by integrating the linearised equations in time.

### *2.3. Fourier formulations for planar and axisymmetric base flows*

For the two-dimensional formulations with a prescribed spanwise or azimuthal wavenumber, the base flow has zero spanwise velocity in planar configurations or zero swirl in axisymmetric configurations.

To avoid complex arithmetic during time integration, the perturbations are represented using a half-mode formulation,

$$\mathbf{q}'(x,y,s,t) = \begin{bmatrix} \tilde{\rho}(x,y,t)cos(\phi) \\ \tilde{u}(x,y,t)cos(\phi) \\ \tilde{v}(x,y,t)cos(\phi) \\ \tilde{w}(x,y,t)sin(\phi) \\ \tilde{T}(x,y,t)cos(\phi) \end{bmatrix}, \tag{15}$$

where $s = z$, $\phi = \beta z$ for planar configurations and $s = \theta$, $\phi = \beta_{\text{axis}}\theta$ for axisymmetric configurations. Here, $\beta \in \mathbb{R}$ is the spanwise angular wavenumber and $\beta_{\text{axis}} \in \mathbb{Z}$ is the azimuthal wavenumber.

Substitution of Eq. (15) into the linearised equations gives a system for $\tilde{\mathbf{q}}(x, y, t) = (\tilde{\rho}, \tilde{u}, \tilde{v}, \tilde{w}, \tilde{T})$ at the prescribed wavenumber. After spatial discretisation, the system has the form of Eq. (12), with a real-valued operator **A** acting on the half-mode amplitudes. Following a derivation analogous to that for the fully three-dimensional formulation yields an eigenvalue problem of the same form as Eq. (14). The zero-wavenumber form recovers the two-dimensional global linear stability analysis problem.

## 3. Numerical methods

This section describes the spatial and temporal discretisation of the nonlinear and linearised governing equations on structured curvilinear grids. The physical coordinates ($x$, $y$, $z$) are mapped to computational coordinates ($\xi$, $\eta$, $\zeta$). Physical-space derivatives are evaluated using the chain rule and the corresponding grid metric terms.

### *3.1. Convective flux discretisation*

The convective terms are discretised using flux vector splitting and upwind reconstruction. MR-WENO reconstruction is used for nonlinear shock-containing flows, whereas linear upwind reconstruction is used for the linearised equations. OpenGLSA supports the Steger-Warming flux splitting [18] with a smooth entropy correction and local Lax-Friedrichs-type flux splitting [19,20]. For the nonlinear equations, characteristic reconstruction can be used to improve robustness in flows containing shocks. For the linearised equations, the split fluxes are reconstructed component-wise in conservative variables.

#### *3.1.1. Flux vector splitting scheme for linearised convective flux*

Following Steger and Warming [18], the linearised convective fluxes are split according to the characteristic speeds of the base flow. Let $\mathbf{U} = (\rho, \rho u, \rho v, \rho w, \rho E)^{\mathrm{T}}$ denote the conservative state vector and $\mathbf{U}' = (\rho', (\rho u)', (\rho v)', (\rho w)', (\rho E)')^{\mathrm{T}}$ its perturbation. The perturbation flux in the physical coordinate direction $x_i$ is written as

$$\mathbf{F}_i' = \mathbf{A}_i'\mathbf{U}', \qquad \mathbf{A}_i' = \frac{\partial \mathbf{F}_i'}{\partial \mathbf{U}'}. \tag{16}$$

The eigenvalues of $\mathbf{A}_i'$ are

$$\lambda_{i,1}' = \lambda_{i,2}' = \lambda_{i,3}' = \bar{u}_i, \qquad \lambda_{i,4}' = \bar{u}_i + \bar{c}, \qquad \lambda_{i,5}' = \bar{u}_i - \bar{c}\,. \tag{17}$$

Here, the overbar denotes the base flow quantities. These eigenvalues are characteristic speeds describing the convection of entropy and vorticity disturbances and the propagation of acoustic disturbances.

Splitting the perturbation flux according to these speeds allows oppositely biased stencils to be used for the positive and negative contributions. The flux Jacobian admits the eigen decomposition

$$\mathbf{A}_i' = \mathbf{R}_i'\mathbf{\Lambda}_i'\mathbf{L}_i', \tag{18}$$

where $\mathbf{R}_i'$ and $\mathbf{L}_i' = (\mathbf{R}_i')^{-1}$ are the right and left eigenvector matrices, respectively. The split perturbation fluxes are

$$(\mathbf{F}_i')^{\pm} = \mathbf{R}_i'(\mathbf{\Lambda}_i')^{\pm}\mathbf{L}_i'\mathbf{U}'. \tag{19}$$

The diagonal entries of $(\mathbf{\Lambda}_i')^{\pm}$ are evaluated using the same smooth entropy correction as for the nonlinear equations:

$$\lambda_{i,r}'^{\pm} = \frac{1}{2}(\lambda_{i,r}' \pm \sqrt{\lambda_{i,r}'^2 + \delta^2}), \tag{20}$$

where $\delta$ is the smoothing parameter. In computational space, the same procedure is applied to the corresponding metric-weighted flux Jacobians.

### *3.1.2. Multi-resolution WENO scheme*

Accurate steady base flows are required for global stability analysis. However, classical WENO schemes, such as WENO-JS [21], may exhibit small-amplitude post-shock oscillations that cause the steady-state residual to hang at the truncation-error level rather than converge to machine zero [22]. OpenGLSA therefore uses the fifth-order multi-resolution WENO (MR-WENO) scheme of Zhu and Shu [16], whose steady-state convergence has been assessed in [23].

For clarity, the reconstruction is described for a scalar conservation law. Its semi-discrete finite-difference form is

$$\frac{du_i(t)}{dt} + \frac{1}{\Delta x}(\hat{f}_{i+1/2} - \hat{f}_{i-1/2}) = 0. \tag{21}$$

Here $u_i$ denotes the solution at the $i$-th grid point and $\hat{f}_{i+1/2}$ denotes the numerical flux at the interface $x_{i+1/2}$. After flux vector splitting, the numerical flux is written as

$$\hat{f}_{i+1/2} = \hat{f}_{i+1/2}^{+} + \hat{f}_{i+1/2}^{-}. \tag{22}$$

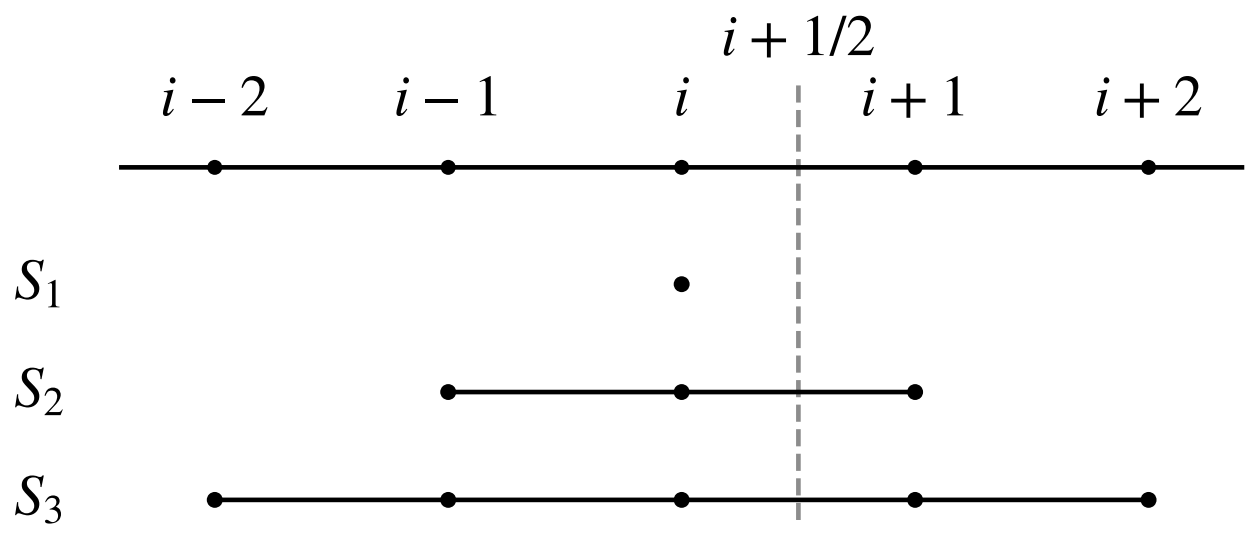

Figure 1 Stencils used by fifth-order MR-WENO.

Unlike the fifth-order WENO-JS scheme, which uses three three-point stencils, the fifth-order MR-WENO scheme employs three nested central stencils,

$$S_1 = \{i\}, \qquad S_2 = \{i-1, i, i+1\}, \qquad S_3 = \{i-2, i-1, i, i+1, i+2\}, \tag{23}$$

as illustrated in Figure 1. Polynomial approximations of degree zero, two and four are constructed on $S_1$, $S_2$ and $S_3$, respectively. Their values at $x_{i+1/2}$ are

$$\varphi_1 = f_i, \tag{24}$$

$$\varphi_2 = -\frac{1}{6} f_{i-1} + \frac{5}{6} f_i + \frac{1}{3} f_{i+1}, \tag{25}$$

$$\varphi_3 = \frac{1}{30} f_{i-2} - \frac{13}{60} f_{i-1} + \frac{47}{60} f_i + \frac{27}{60} f_{i+1} - \frac{1}{20} f_{i+2}. \tag{26}$$

Here, $f_i$ denotes the positive split flux value at grid point $i$ and the superscript $(*)^+$ is omitted for brevity.

Then the three candidate polynomials can be expressed in hierarchical form as

$$\tilde{\varphi}_1 = \varphi_1, \tag{27}$$

$$\tilde{\varphi}_2 = \frac{1}{d_{2,2}} \varphi_2 - \frac{d_{1,2}}{d_{2,2}} \tilde{\varphi}_1, \tag{28}$$

$$\tilde{\varphi}_3 = \frac{1}{d_{3,3}} \varphi_3 - \frac{d_{1,3}}{d_{3,3}} \tilde{\varphi}_1 - \frac{d_{2,3}}{d_{3,3}} \tilde{\varphi}_2. \tag{29}$$

The linear weights used are

$$(d_{1,2}, d_{2,2}) = (\frac{1}{11}, \frac{10}{11}) \qquad (d_{1,3}, d_{2,3}, d_{3,3}) = (\frac{1}{111}, \frac{10}{111}, \frac{100}{111}), \tag{30}$$

and they ensure that the hierarchical construction gives

$$\sum_{l=1}^{3} d_{l,3} \tilde{\varphi}_l = \varphi_3 . \tag{31}$$

The nonlinear weights are computed from smoothness indicators associated with the three stencils. Since stencil $S_1$ has only one point, its smoothness indicator is constructed differently from those on $S_2$ and $S_3$. First, define

$$s_0 = (f_i - f_{i-1})^2, \qquad s_1 = (f_{i+1} - f_i)^2 , \tag{32}$$

$$(g_{01}, g_{11}) = \begin{cases} (\frac{1}{11}, \frac{10}{11}), & s_0 \ge s_1, \\ (\frac{10}{11}, \frac{1}{11}), & s_0 < s_1, \end{cases} \tag{33}$$

and

$$\tilde{s}_0 = g_{01}[1 + \frac{(s_0 - s_1)^2}{s_0 + \varepsilon}], \qquad \tilde{s}_1 = g_{11}[1 + \frac{(s_0 - s_1)^2}{s_1 + \varepsilon}], \qquad s = \tilde{s}_0 + \tilde{s}_1 , \tag{34}$$

then the first smoothness indicator is given as

$$IS_1 = [\frac{\tilde{s}_0 (f_i - f_{i-1}) + \tilde{s}_1 (f_{i+1} - f_i)}{s}]^2 . \tag{35}$$

The remaining two smoothness indicators are computed as follows:

$$IS_2 = \frac{13}{12}(f_{i-1} - 2f_i + f_{i+1})^2 + \frac{1}{4}(f_{i-1} - f_{i+1})^2,$$

$$\begin{aligned} IS_3 &= \frac{1}{144}(f_{i-2} - 8f_{i-1} + 8f_{i+1} - f_{i+2})^2 \\ &+ \frac{1}{15600}(-11f_{i-2} + 174f_{i-1} - 326f_i + 174f_{i+1} - 11f_{i+2})^2 \\ &+ \frac{781}{2880}(-f_{i-2} + 2f_{i-1} - 2f_{i+1} + f_{i+2})^2 \\ &+ \frac{1421461}{1310400}(f_{i-2} - 4f_{i-1} + 6f_i - 4f_{i+1} + f_{i+2})^2. \end{aligned} \tag{36}$$

The nonlinear weights are then computed as

$$\alpha_{l,3} = \frac{\tilde{\alpha}_{l,3}}{\sum_{l=1}^{3} \tilde{\alpha}_{l,3}}, \qquad \tilde{\alpha}_{l,3} = d_{l,3}(1 + \frac{\tau_3}{\varepsilon + IS_l}), \quad l = 1,2,3, \tag{37}$$

where $\tau_3$ is defined as

$$\tau_3 = (\frac{|IS_3 - IS_1| + |IS_3 - IS_2|}{2})^2. \tag{38}$$

Here $\varepsilon = 10^{-6}$ is a small positive number to prevent zero denominators in Eq. (34) and Eq. (37) .

Finally, the reconstructed positive numerical flux is

$$\hat{f}^{+}_{i+1/2} = \sum_{l=1}^{3} \alpha_{l,3} \tilde{\varphi}_l . \tag{39}$$

The negative split flux is reconstructed using the same procedure on the corresponding mirror-symmetric stencils.

The MR-WENO reconstruction is used in shock-containing flows, while the linear upwind reconstruction $\varphi_3$ in Eq. (26) can be used for smooth flows. For the linearised equations, the split perturbation fluxes are reconstructed by linear upwind reconstruction to ensure the propagator remains linear. Applying MR-WENO reconstruction to the perturbation field may also introduce spurious oscillations in regions where the perturbations should decay to zero. A comparison of the two reconstruction schemes for the perturbation fluxes is presented in Sec.5.5. Similar spurious oscillations near shocks were reported in [24] when WENO reconstruction was used for resolvent analysis of shock-laden flows.

#### *3.1.3. Positivity-preserving flux limiter*

High-order WENO reconstruction may produce negative density or pressure in strong-shock or low-density flows, causing the failure of the base-flow computation. To improve robustness, the positivity-preserving limiter of Zhang and Shu [20] is optionally applied to the nonlinear convective terms. It is used with the local Lax-Friedrichs splitting, with the maximum of characteristic speeds over the whole stencil. After characteristic reconstruction, the split fluxes are projected back to conservative variables and limited at every SSP Runge–Kutta stage. A density correction is followed by a pressure correction, which scales each reconstructed split flux towards its upwind point value. The density threshold is taken as $\min(10^{-13}, \rho_{\min})$, where $\rho_{\min}$ is the global minimum split-state density. The limited positive and negative split fluxes are combined into a single interface flux, so conservation is preserved. The limiter is not applied to the linearised equations.

### *3.2. Central schemes for viscous flux and metric terms*

OpenGLSA supports fourth- and sixth-order central schemes for viscous fluxes and grid metrics. The viscous terms are evaluated in the fully expanded form, with second derivatives discretised directly to alleviate odd-even decoupling [25]. The sixth-order central scheme is used for the cases presented here. At interior points, the first and second derivatives for a grid spacing of $\Delta\xi$ are

$$f_i' = \frac{(f_{i+3} - f_{i-3}) - 9(f_{i+2} - f_{i-2}) + 45(f_{i+1} - f_{i-1})}{60\Delta\xi}, \tag{40}$$

$$f_i'' = \frac{2(f_{i+3} + f_{i-3}) - 27(f_{i+2} + f_{i-2}) + 270(f_{i+1} + f_{i-1}) - 490 f_i}{180\Delta\xi^2}. \tag{41}$$

At physical boundaries requiring one-sided closures, fourth-order one-sided approximations are used for first derivatives, while third-order approximations are used for second derivatives at the two points nearest each boundary. Fourth-order central approximations are used where the sixth-order interior stencil is unavailable. The boundary formulas are described in Lusher et al. [26].

### *3.3. Explicit Runge-Kutta time integration*

Both the nonlinear and linearised equations are advanced using the three-stage, third-order SSP Runge-Kutta scheme [21]. For a semi-discrete system

$$\frac{d\mathbf{U}}{dt} = R(\mathbf{U}), \tag{42}$$

the Runge-Kutta stages are

$$\begin{cases} \mathbf{U}^{(1)} = \mathbf{U}^n + \Delta t R(\mathbf{U}^n), \\ \mathbf{U}^{(2)} = \dfrac{3}{4}\mathbf{U}^n + \dfrac{1}{4}\mathbf{U}^{(1)} + \dfrac{1}{4}\Delta t R(\mathbf{U}^{(1)}), \\ \mathbf{U}^{n+1} = \dfrac{1}{3}\mathbf{U}^n + \dfrac{2}{3}\mathbf{U}^{(2)} + \dfrac{2}{3}\Delta t R(\mathbf{U}^{(2)}), \end{cases} \tag{43}$$

where $R(\mathbf{U})$ is the residual of the nonlinear or linearised equations.

### *3.4. Axis treatment for axisymmetric configurations*

For two-dimensional axisymmetric configurations, the symmetry axis is aligned with the $x$ axis, with $y = 0$ on the axis. The terms containing $1/y$ in Eqs. (8)-(10) are formally singular. The axis treatment proposed by Sandberg [27] is adopted. Ghost-point values across the axis are assigned according to the appropriate parity conditions, which depend on both the flow variable and the prescribed azimuthal wavenumber $\beta_{\text{axis}}$. The axis values of these terms are obtained by interpolation from the four points above the axis, and the resulting contributions are included in the residual.

## 4. Solution procedures and implementation

This section presents the selective frequency damping procedure for base-flow computation, the time-stepping Arnoldi method for global stability analysis, and the software organisation.

### *4.1. Steady base flow computation using selective frequency damping*

For base flows that are unstable within the selected computational domain, steady solutions are computed using the encapsulated formulation of selective frequency damping (SFD) proposed by Jordi et al. [28], based on the original method of Åkervik et al. [29]. This formulation combines the existing time-stepping solver with a separate feedback damping and temporal filtering update stage. The nonlinear solver first advances the flow state through one complete Runge-Kutta time step to obtain an auxiliary flow state $\mathbf{U}^*$. The flow state $\mathbf{U}^{n+1}$ and corresponding filtered solution $\mathbf{U}_f$ are then updated as

$$\begin{pmatrix} \mathbf{U}^{n+1} \\ \mathbf{U}_f^{n+1} \end{pmatrix} = \exp(\mathcal{L}\Delta t) \begin{pmatrix} \mathbf{U}^* \\ \mathbf{U}_f^n \end{pmatrix}, \qquad \mathcal{L} = \begin{pmatrix} -\chi\mathbf{I} & \chi\mathbf{I} \\ \mathbf{I}/\Delta & -\mathbf{I}/\Delta \end{pmatrix}, \tag{44}$$

where $\chi$ is the control coefficient, $\Delta$ is the filter width, and $\mathbf{I}$ is the identity matrix. Convergence depends on the choice of $\chi$ and $\Delta$, whose values are problem dependent. At convergence, the flow state and filtered state coincide, the feedback term vanishes, and a steady solution of the original governing equations is recovered.

### *4.2. Global stability analysis with Matrix-free time-stepping*

Explicit assembly and factorisation of the Jacobian matrix can require substantial memory. OpenGLSA therefore adopts a matrix-free time-stepping approach for global stability analysis [2]. For a steady base flow, the linearised system admits the formal solution

$$\mathbf{q}'(\Delta T) = \exp(\mathbf{A}\Delta T)\mathbf{q}'(0) = \mathbf{G}_{\Delta T}\mathbf{q}'(0), \tag{45}$$

where $\mathbf{G}_{\Delta T}$ is the propagator over the integration interval $\Delta T$. The action of the propagator on a vector is evaluated by integrating the linearised equations over $\Delta T$, without explicitly assembling either $\mathbf{A}$ or $\mathbf{G}_{\Delta T}$.

Starting from a random initial perturbation vector $\mathbf{v}_1$ of unit norm, successive applications of the propagator generate the Krylov subspace

$$\mathcal{K}_m(\mathbf{G}_{\Delta T}, \mathbf{v}_1) = \mathrm{span}\{\mathbf{v}_1, \mathbf{G}_{\Delta T}\mathbf{v}_1, \ldots, \mathbf{G}_{\Delta T}^{m-1}\mathbf{v}_1\}. \tag{46}$$

The Arnoldi iteration constructs an orthonormal basis of this subspace, stored as the columns of $\mathbf{V}_m$, and projects the propagator onto a small upper Hessenberg matrix,

$$\mathbf{H}_m = \mathbf{V}_m^* \mathbf{G}_{\Delta T} \mathbf{V}_m, \tag{47}$$

where the superscript * denotes the conjugate transpose. The eigenpairs ($\sigma_k$, $\mathbf{y}_k$) of $\mathbf{H}_m$ provide Ritz approximations to the eigenpairs of the propagator. The corresponding global modes and temporal eigenvalues are recovered as

$$\hat{\mathbf{q}}_k \approx \mathbf{V}_m \mathbf{y}_k, \qquad \lambda_k = \frac{\ln(\sigma_k)}{\Delta T}. \tag{48}$$

OpenGLSA uses PARPACK, the parallel implementation of ARPACK [17], to perform implicitly restarted Arnoldi iteration. The library is taken from ARPACK-NG, a community-maintained distribution of ARPACK.

### *4.3. Code organisation*

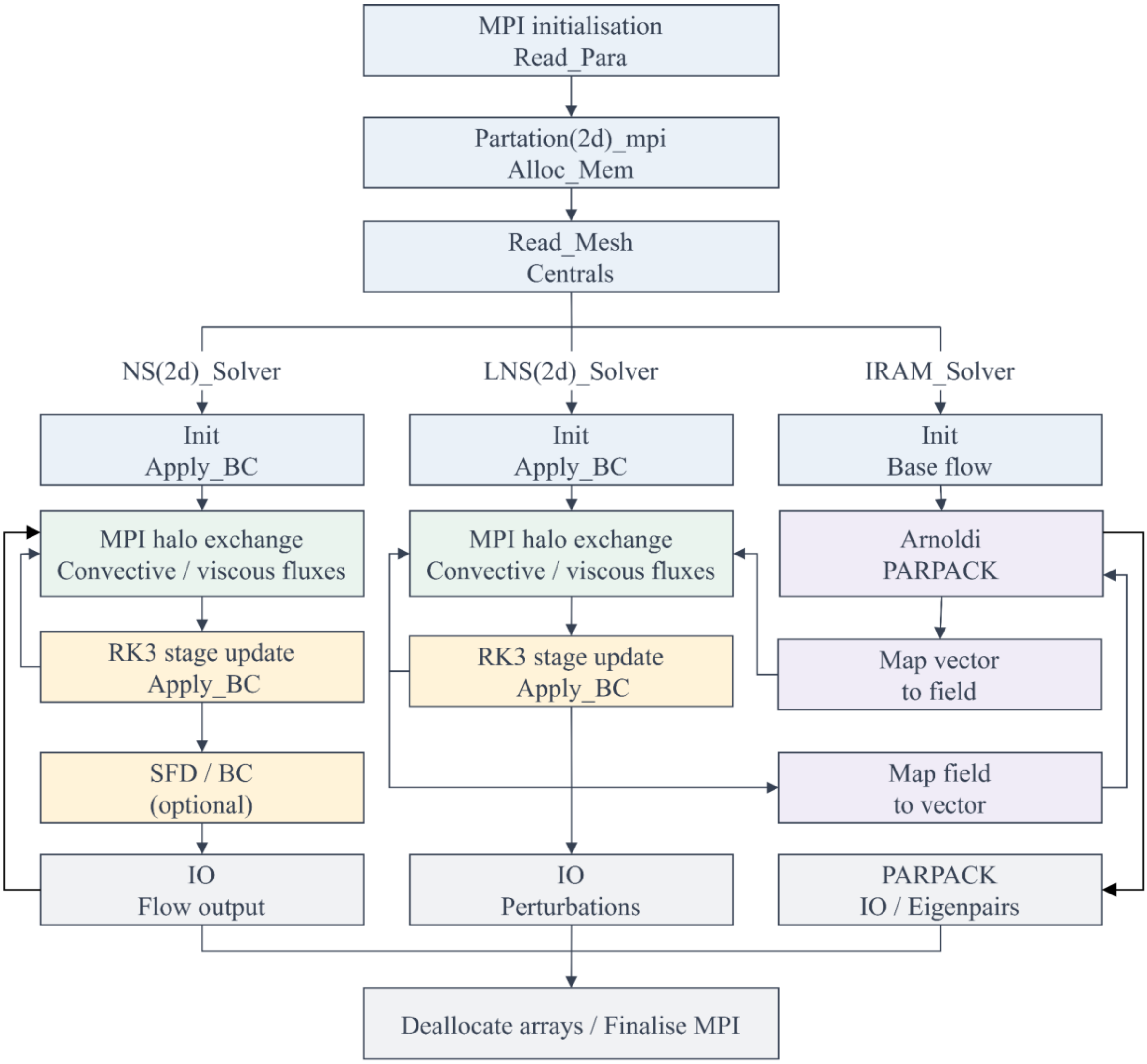

Figure 2 Code organisation and computational workflow of OpenGLSA.

Figure 2 shows the code organisation and computational workflow of OpenGLSA. The program provides three run modes, nonlinear flow simulation, linear perturbation evolution and global stability analysis. The two- and three-dimensional implementations follow the same software organisation. The three modes share the same grid, metric, halo-exchange and time-integration routines, while the nonlinear and linearised equations use separate flux and boundary condition routines. The code organisation, MPI parallelisation and parts of the nonlinear mode are adapted from OpenCFD-SC, part of the OpenCFD software suite [30,31].

For global stability analysis, the Arnoldi solver couples the linearised solver with PARPACK through its reverse-communication interface. Each requested operator action is evaluated by copying the Arnoldi vector into the local perturbation fields, integrating the linearised equations over $\Delta T$, and returning the resulting vector to PARPACK.

New configurations are defined by implementing a pair of user routines in ***BoundaryConditions.f90***, one for the nonlinear boundary conditions and one for the linearised counterparts. The two must be consistent, since the perturbation boundary conditions directly affect the computed eigenvalues. The routines for the validation cases in Section 5 can be used as examples.

The main input and output files are described below:

***control.nml***: specifies the flow parameters, grid dimensions, boundary condition settings and parameters for implicitly restarted Arnoldi iteration.

***grid(3d).dat***: contains the computational grid. An auxiliary tool is provided to convert PLOT3D grids to this format. The grid index orientation must be consistent with the metric calculation and boundary condition settings.

***flow(3d)-xxx.dat***: contains density, velocity components and temperature from the nonlinear solver, saved at a user-specified time step interval. A steady flow solution must be renamed ***flow(3d).dat*** to serve as the base flow required by the linearised and Arnoldi solvers.

***fluc(3d)-xxx.dat***: contains perturbations of density, all three velocity components and temperature. These files are generated in linear perturbation evolution mode.

***Eigenvalues.dat***: contains the converged eigenvalues obtained by the Arnoldi solver.

***Mode-xxx.dat***: contains the eigenmodes corresponding to the eigenvalues in ***Eigenvalues.dat***.

Compilation instructions, parameter definitions and example cases are provided in the project repository.

## 5. Validation and results

Five benchmark cases are selected to assess the main capabilities of OpenGLSA. For each case, the base flow and global stability results are compared with published data: a cylinder flow [14] and a sphere flow [32] at low Mach number, a square [33,34] and a cubic [35,36] lid-driven cavity flow, and a hypersonic compression ramp [5,6]. All computations were performed using 80 MPI processes on a server equipped with four Intel Xeon Gold 6148 processors operating at 2.40 GHz, providing 80 physical cores in total and approximately 251 GiB of system memory. The code was compiled using compilers and libraries bundled in Intel OneAPI 2021.

### *5.1. Cylinder flow*

#### *5.1.1. Base flow*

Flow past a circular cylinder is considered at $Ma_\infty = 0.1$ and $Re_\infty = 100$. The Reynolds number is based on the cylinder diameter $D = 1$ and the freestream temperature is set to $T_\infty = 288.15$ K. The computational domain is a single-block O-grid with an outer radius of $100D$, comprising 316 points in the wall-tangent direction and 301 points in wall-normal direction.

An adiabatic no-slip wall boundary condition and a non-reflecting far-field boundary condition are prescribed on the surface of the cylinder and outer boundary, respectively. Periodicity is imposed across the branch cut of the O-grid. The perturbation boundary conditions are obtained by linearising the corresponding boundary conditions used for the base flow. Linear reconstruction is used for the convective fluxes to reduce computational cost, as the flow is smooth everywhere.

The selected Reynolds number exceeds the critical value of approximately 46.8 [37] for the first bifurcation in two-dimensional cylinder flow. Hence, the SFD method is used to obtain the steady base flow. The control coefficient $\chi$ and filter width $\Delta$ are set to 0.4273 and 3.2650, respectively, close to the values used in [38]. The convergence history of residuals is shown in Figure 3. The two average residuals are defined as

$$\text{Res} = \sum_{i=1}^{N} \frac{\sum_{k=1}^{4} | U_{k,i}^{n+1} - U_{k,i}^{n} |}{4 \times N \times \Delta t}, \qquad \text{Res}_{\text{SFD}} = \sum_{i=1}^{N} \frac{\sum_{k=1}^{4} | U_{k,i}^{n} - U_{f,k,i}^{n} |}{4 \times N}, \tag{49}$$

where $N$ is the total number of grid points. The residuals eventually reach levels of approximately $10^{-10}$.

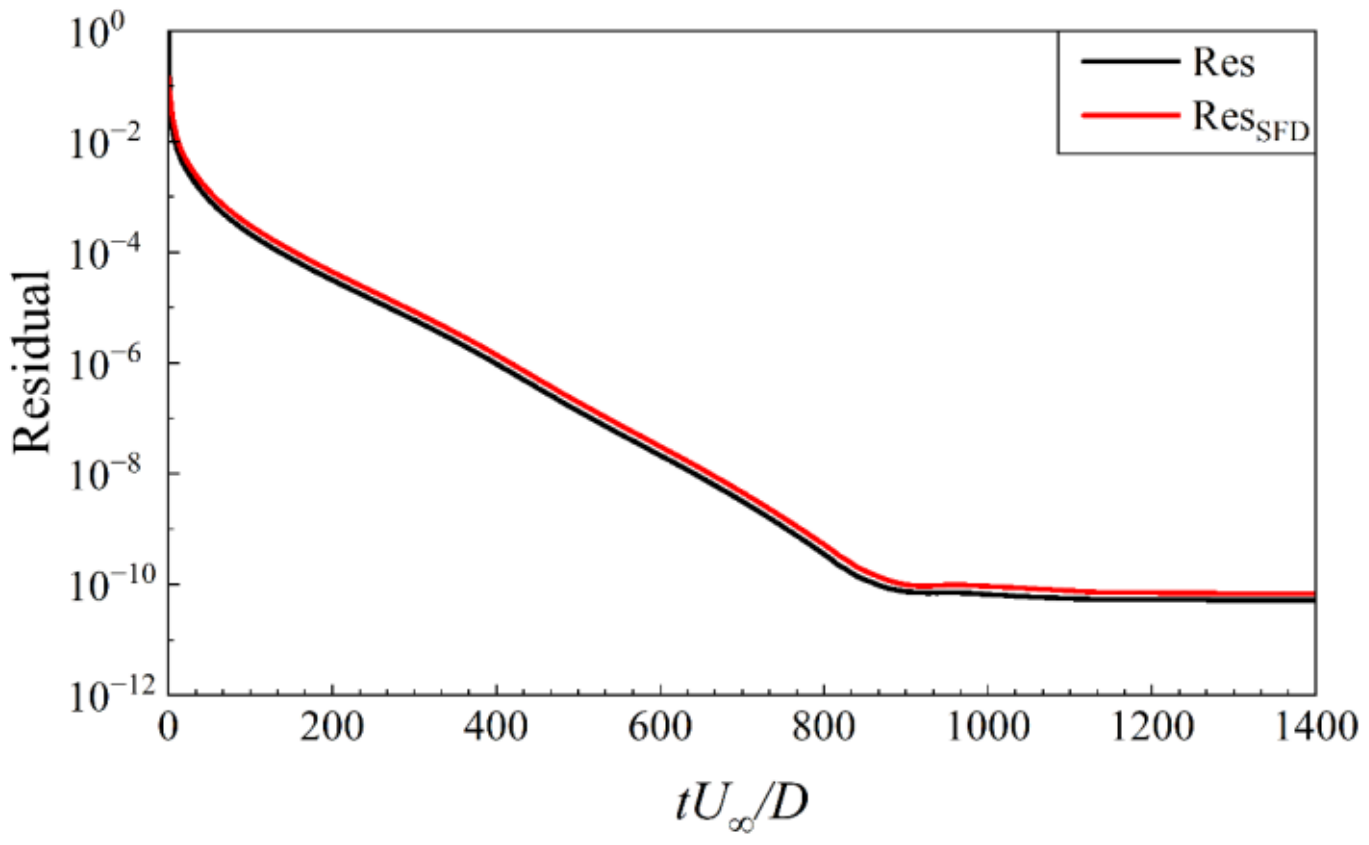

Figure 3 Convergence history of residuals for the cylinder base flow.

The vorticity field of the converged base flow is shown in Figure 4, which reveals a pair of separation bubbles arranged symmetrically about the wake centreline downstream of the circular cylinder.

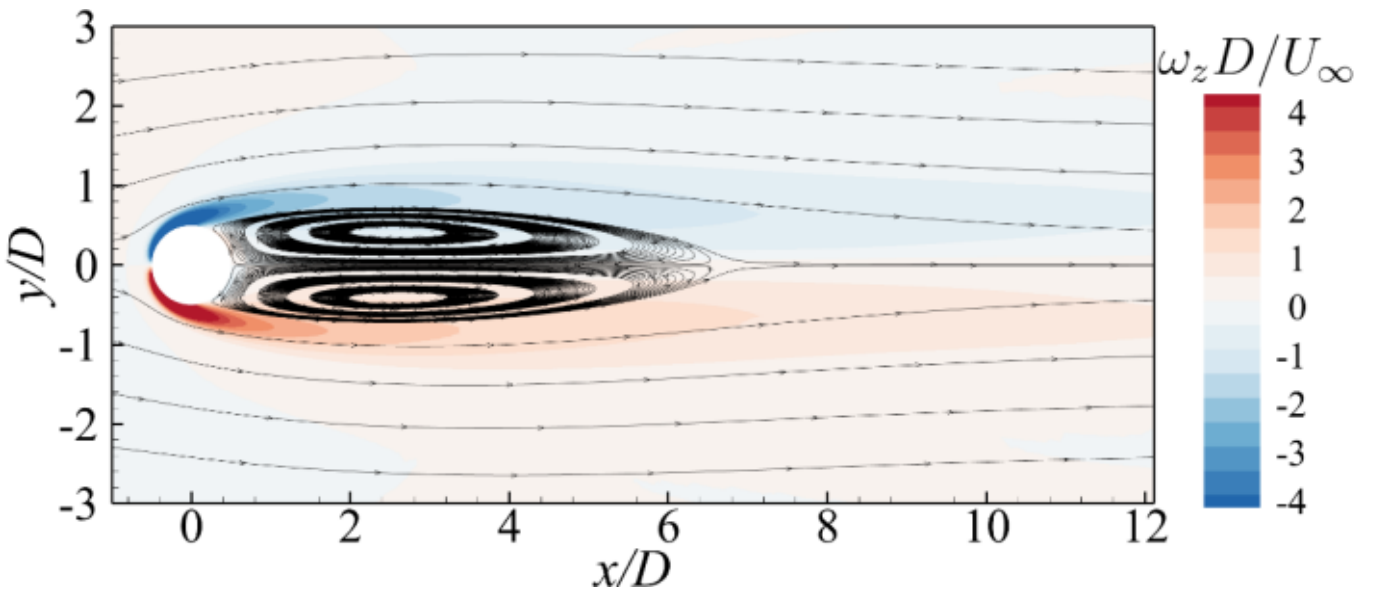

Figure 4 Vorticity of the cylinder base flow at $Ma_\infty = 0.1$, $Re_\infty = 100$.

### *5.1.2. Global linear modes*

Global stability analysis is performed over a range of spanwise angular wavenumbers from 0 to 0.9, and the results are compared with data digitised from figures in [14]. The propagation interval is set to $\Delta T = 1$, with a time-integration step of $\Delta t = 2.5\times10^{-4}$. The Krylov subspace dimension is set to 20, and the maximum number of restarts is 50. The relative convergence tolerance for the Ritz values is set to $10^{-8}$.

For $\beta = 0$, the unstable complex-conjugate pair converged within 9 Arnoldi cycles, requiring 164 propagator evaluations and approximately 17.5 min on 80 cores. The peak memory, summed over all MPI processes as recorded by the Slurm accounting system, was approximately 3.58 GB for OpenGLSA and 43.61 GB for BROADCAST with the same grid, corresponding to a reduction of about 91.8%. This memory reduction comes at the cost of computing time: BROADCAST obtained ten eigenvalues near a prescribed shift in a single shift-invert computation within approximately 8 min on the same number of cores. In contrast, the time-stepping approach requires no prior knowledge of the shift and converges directly to the most amplified modes.

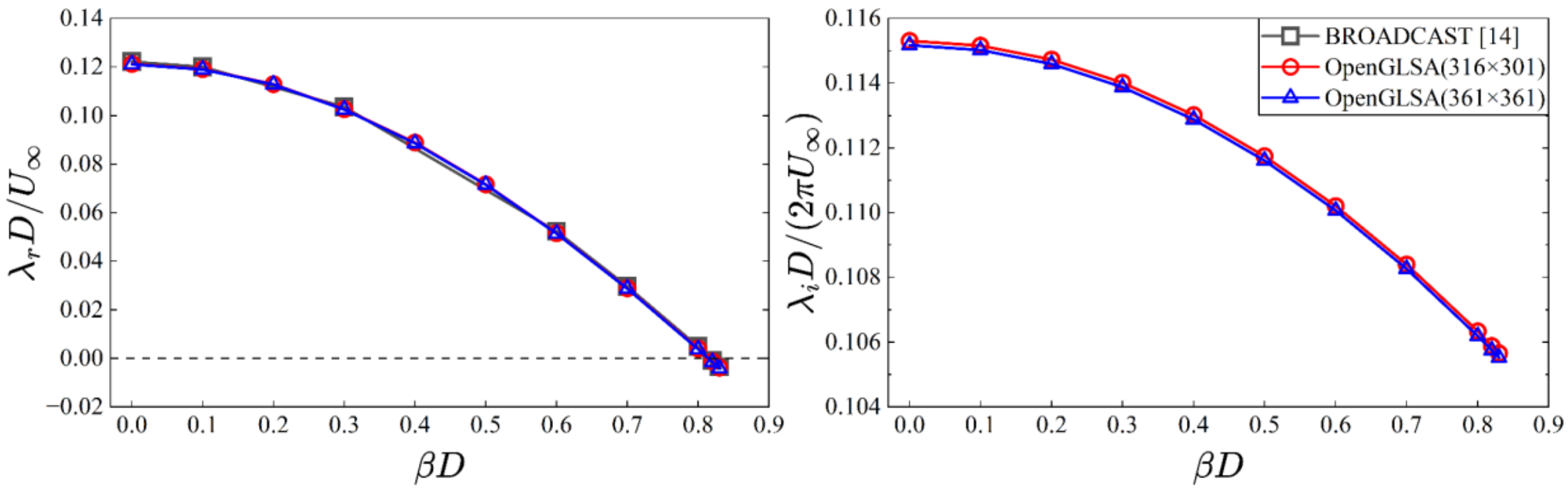

Figure 5 Growth rate and frequency of the vortex-shedding branch as functions of the spanwise angular wavenumber. Comparison between OpenGLSA and BROADCAST [14].

Figure 5 shows the growth rate and frequency of the vortex-shedding mode as functions of the spanwise wavenumber. They both decrease as the wavenumber increases, and the mode becomes stable near $\beta = 0.81$. The results agree well with those obtained from BROADCAST in [14]. Refining the grid from (316×301) to (361×361) produces nearly identical growth rates and only small changes in frequency, indicating that the computed eigenvalues are insensitive to this grid refinement over the wavenumber range examined.

The real parts of the velocity components of the most unstable mode are plotted in Figure 6. The perturbations are concentrated in the downstream wake, with the streamwise component antisymmetric and the transverse component symmetric about the wake centreline. Both components exhibit alternating positive and negative regions along the streamwise direction, consistent with the spatial structure of a vortex-shedding mode.

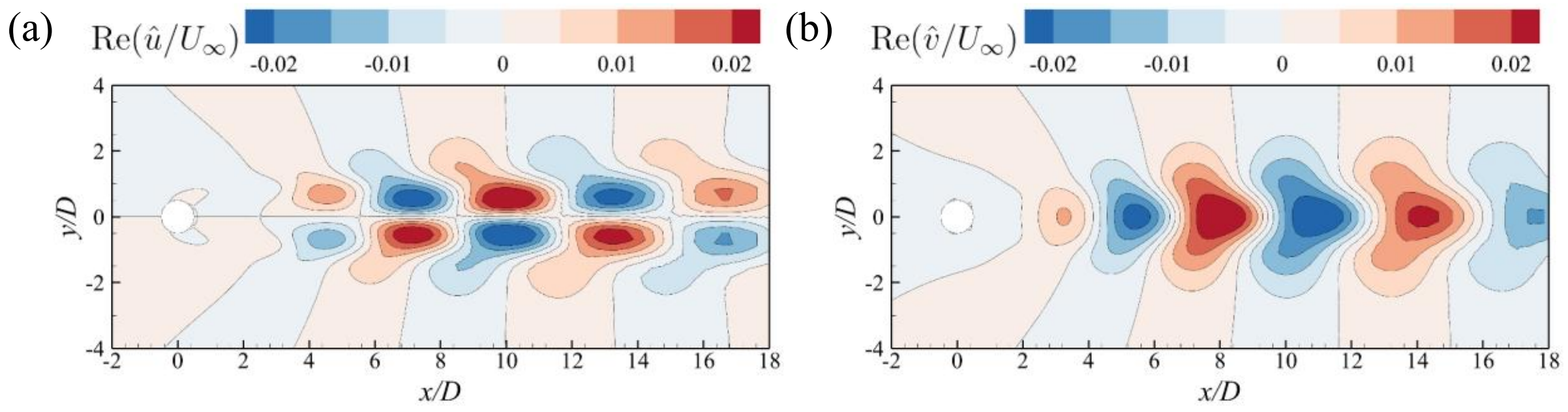

Figure 6 Real part of the leading eigenmode with spanwise angular wavenumber $\beta = 0$: (a) streamwise, (b) transverse velocity perturbations.

### 5.2. Axisymmetric sphere flow

#### 5.2.1. Base flow

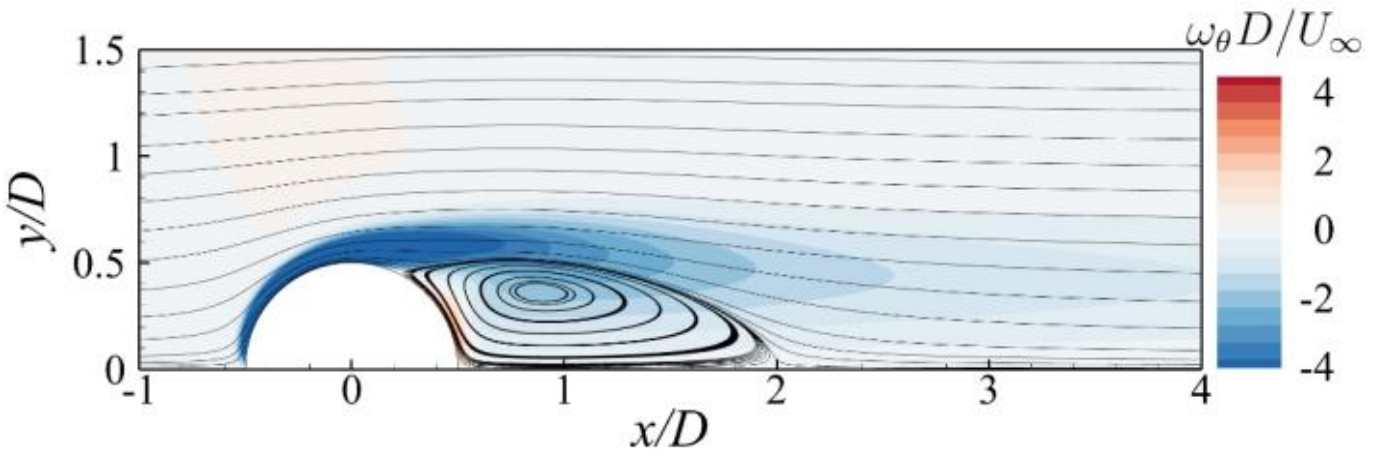

Figure 7 Vorticity of the sphere base flow at $Ma_\infty = 0.1$, $Re_\infty = 213$.

Flow past a sphere is considered to assess the implementation for axisymmetric configurations. The freestream Mach number is set to $Ma_\infty = 0.1$ for comparison with the incompressible simulations. The bifurcation sequence of a sphere has been studied by Schuh Frantz et al. [32]. The first pitchfork bifurcation, occurring at a Reynolds number of approximately 212 [32,39–43], changes the wake symmetry from axisymmetric to planar-symmetric. It can be studied using a two-dimensional axisymmetric base flow coupled with three-dimensional perturbations. The sphere diameter $D$ is used as the reference length.

The freestream temperature is set to $T_\infty = 288.15$ K. The computational domain has the same outer radius as in the cylinder case and covers the upper half-plane, with 241 points in the wall-tangent direction and 361 in the wall-normal direction. An adiabatic no-slip wall boundary condition is imposed on the sphere surface, and a far-field boundary condition is prescribed at the outer semicircular boundary. Axis boundary conditions are imposed on the two remaining edges.

The vorticity field of the sphere base flow at $Re_\infty = 213$ is given in Figure 7. A recirculation region forms behind the sphere, bounded by the separated shear layer, with its extent indicated by the white dash-dotted line.

### *5.2.2. Global linear modes*

Since the pitchfork instability is non-axisymmetric, the axisymmetric base flow remains stable within the axisymmetric formulation and is obtained by direct time integration without SFD. GSA is performed for azimuthal wavenumber $\beta_{\text{axis}} = 1$. The propagation interval is set to $\Delta T = 1$, with a time-integration step of $\Delta t = 1.0\times10^{-4}$. The Krylov subspace dimension is set to 40, and the maximum number of restarts is 20. The relative convergence tolerance for the Ritz values is set to $10^{-8}$.

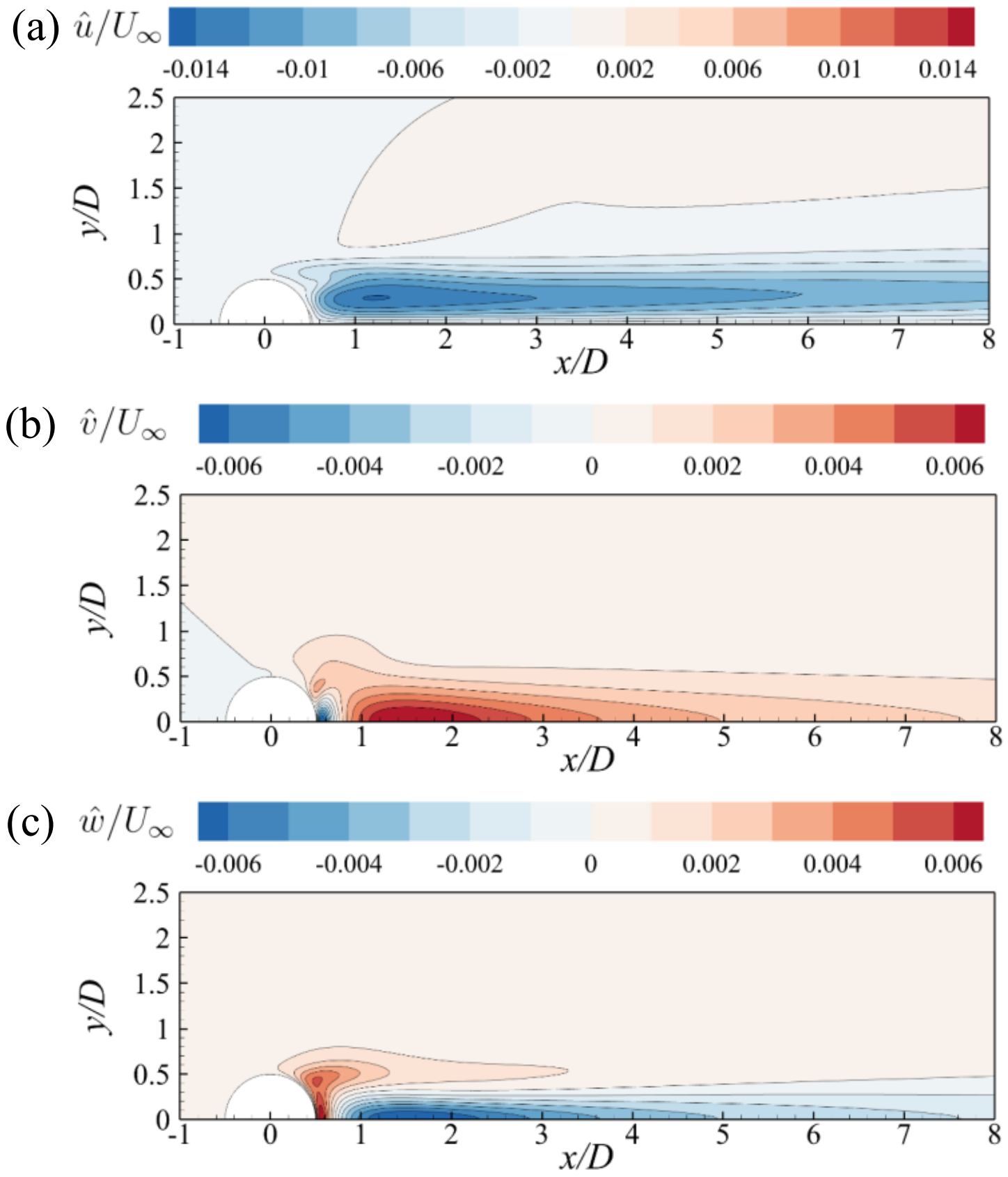

Figure 8 Leading eigenmode with azimuthal wavenumber $\beta_{\text{axis}} = 1$ at $Ma_\infty = 0.1$, $Re_\infty = 213$: (a) axial, (b) radial, and (c) azimuthal velocity perturbations.

Figure 8 shows the leading unstable mode at $Re_\infty = 213$ and $\beta_{\text{axis}} = 1$. The corresponding eigenvalue is real, indicating a non-oscillatory instability. The axial velocity perturbation forms an elongated structure in the downstream wake, whereas the radial and azimuthal components exhibit regions of opposite sign near the rear of the sphere and extend downstream. The azimuthal dependence associated with $\beta_{\text{axis}} = 1$ breaks the rotational symmetry of the axisymmetric base flow. On the axis, the axial perturbation vanishes while the radial and azimuthal components are of equal magnitude and opposite sign, consistent with the regularity conditions for $\beta_{\text{axis}} = 1$ imposed by the axis treatment.

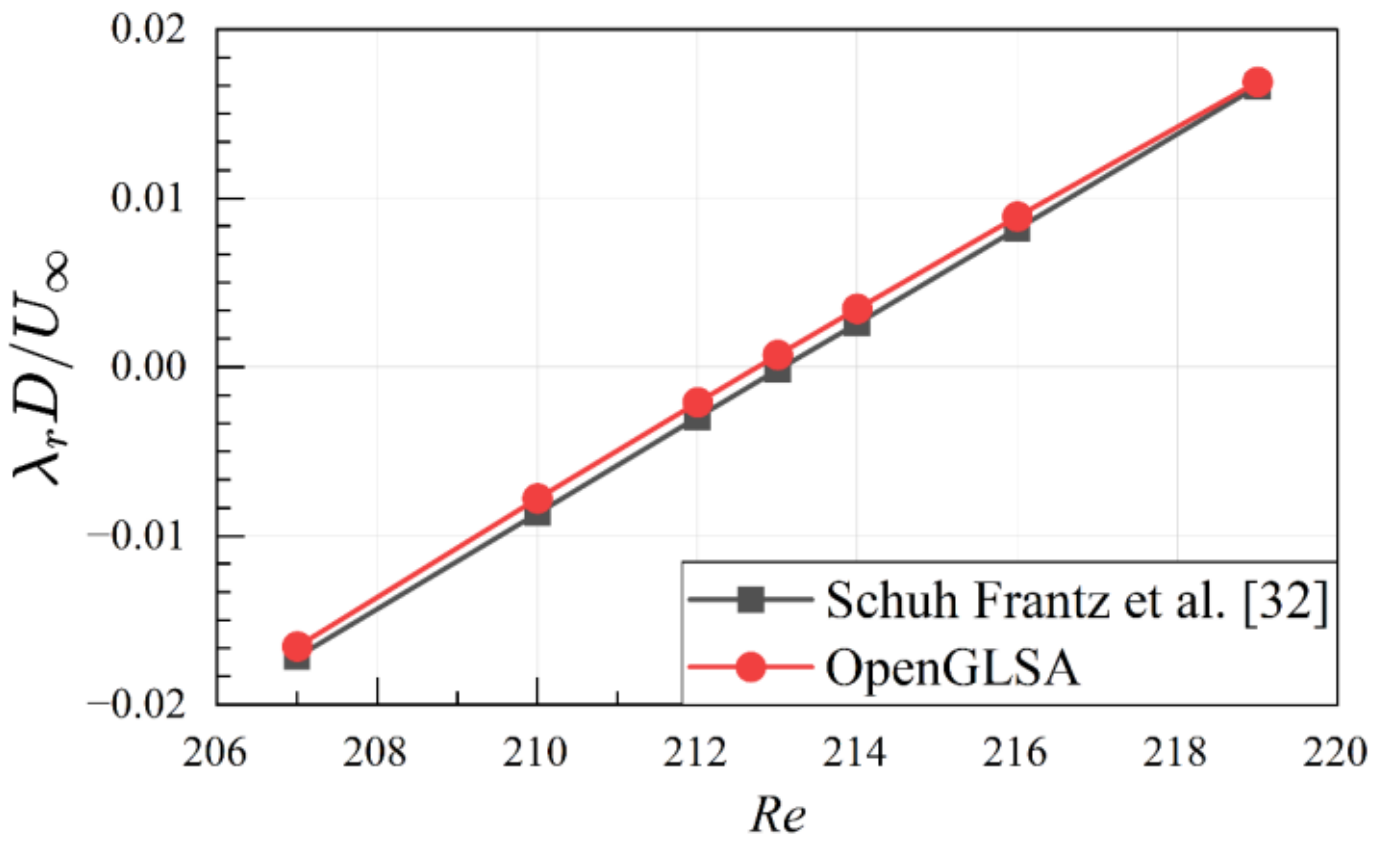

Figure 9 Variation of the growth rate of the leading eigenmode with respect to the Reynolds number. Comparison with Schuh Frantz et al. [32].

Figure 9 compares the growth rates of the leading eigenmode with those reported by Schuh Frantz et al. [32] at several Reynolds numbers near the critical value. Both datasets show a nearly linear increase in growth rate with Reynolds number, with the present values slightly higher than the reference results. The predicted critical Reynolds number obtained by a linear fit is approximately 212.47, which is 0.3% lower than the reported value of 213.11. The fitted slope of 0.00279 is also very close to the reported value of 0.0028, indicating excellent agreement with the literature.

### *5.3. Square lid-driven cavity flow*

#### *5.3.1. Base flow*

The square regularised lid-driven cavity flow is considered to assess the planar formulation for a confined flow. The unit square computational domain is surrounded by three stationary isothermal walls and one moving isothermal wall, with 401 points assigned to each edge. The nondimensional wall temperature is set to 1. The Mach number and Reynolds number based on the edge length $L_{\text{ref}} = 1$ are $Ma_\infty = 0.1$ and $Re_\infty = 900$, respectively. A regularised tangential velocity profile is adopted at the top wall to avoid singularities at the two corners:

$$u(x,1) = [1-(2x-1)^{18}]^2 . \tag{50}$$

The compatibility condition is enforced for the pressure at the walls. All settings are consistent with those in [33].

Bergamo et al. [33] reported that a weak imposition of the isothermal wall boundary condition may fail to enforce zero mass flux through the cavity walls, leading to a persistent density error during steady state convergence. To prevent the drift of total mass, the total mass of the stationary initial flow is recorded, and the density is uniformly scaled by the ratio of the initial total mass to the current total mass, while velocity components and temperature remains the same. With this correction, the residual decreases to the round-off error level and subsequently fluctuates around it, as shown in Figure 10, whereas residual stagnation was reported in [33].

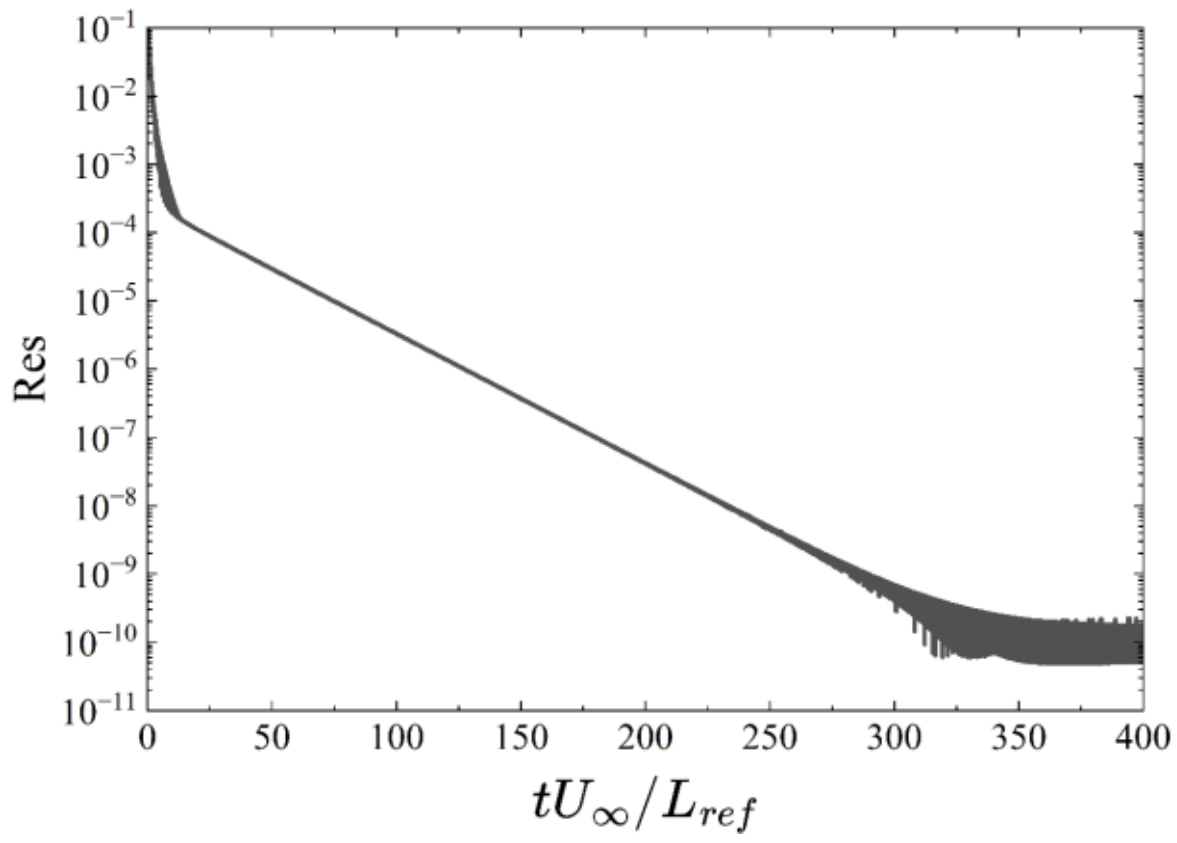

Figure 10 Convergence history of the residual for the square regularised lid-driven cavity base flow.

Figure 11 shows the base flow contours, with a main vortex in the centre and two small vortices in the lower-left and lower-right corners. The velocity contours are also very similar to those reported in [33] for $Re_\infty = 400$ case.

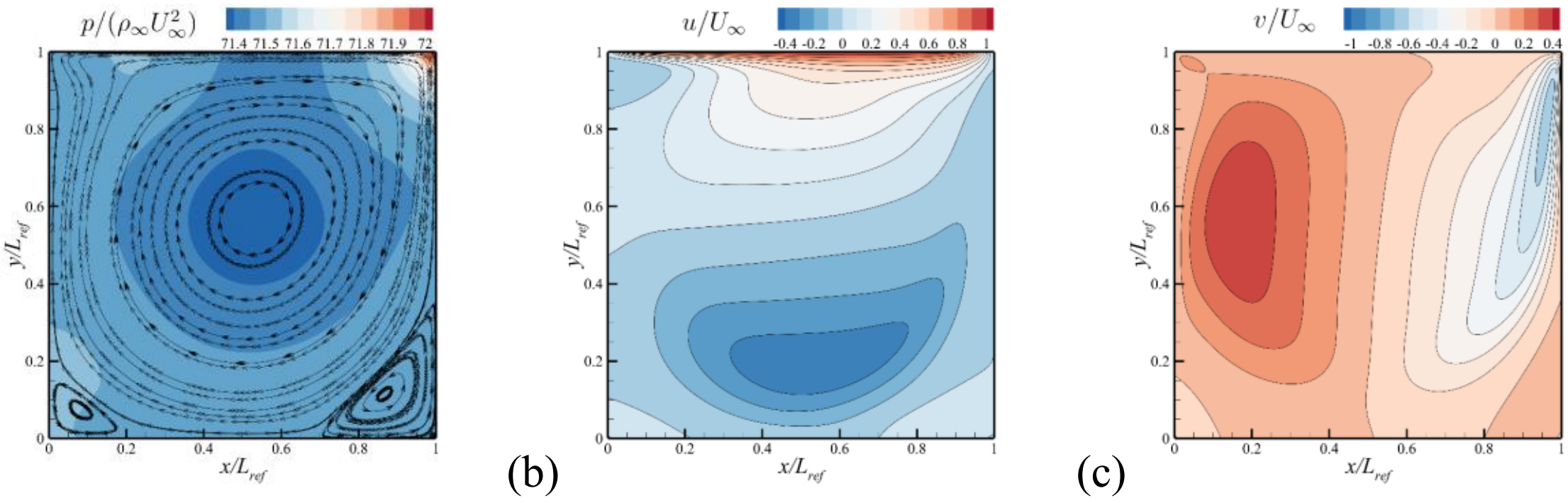

Figure 11 Base flow fields of the square regularised lid-driven cavity. (a) Pressure contours and streamlines, (b) x velocity contours and (c) y velocity contours.

### 5.3.2. Global linear modes

For global stability analysis, the spanwise angular wavenumber is $\beta = 7.35$ and the Krylov subspace dimension is 41. The propagation interval is set to $\Delta T = 0.5$, with a time-integration step of $\Delta t = 1.0\times10^{-4}$. The relative convergence tolerance for the Ritz values is set to $10^{-8}$.

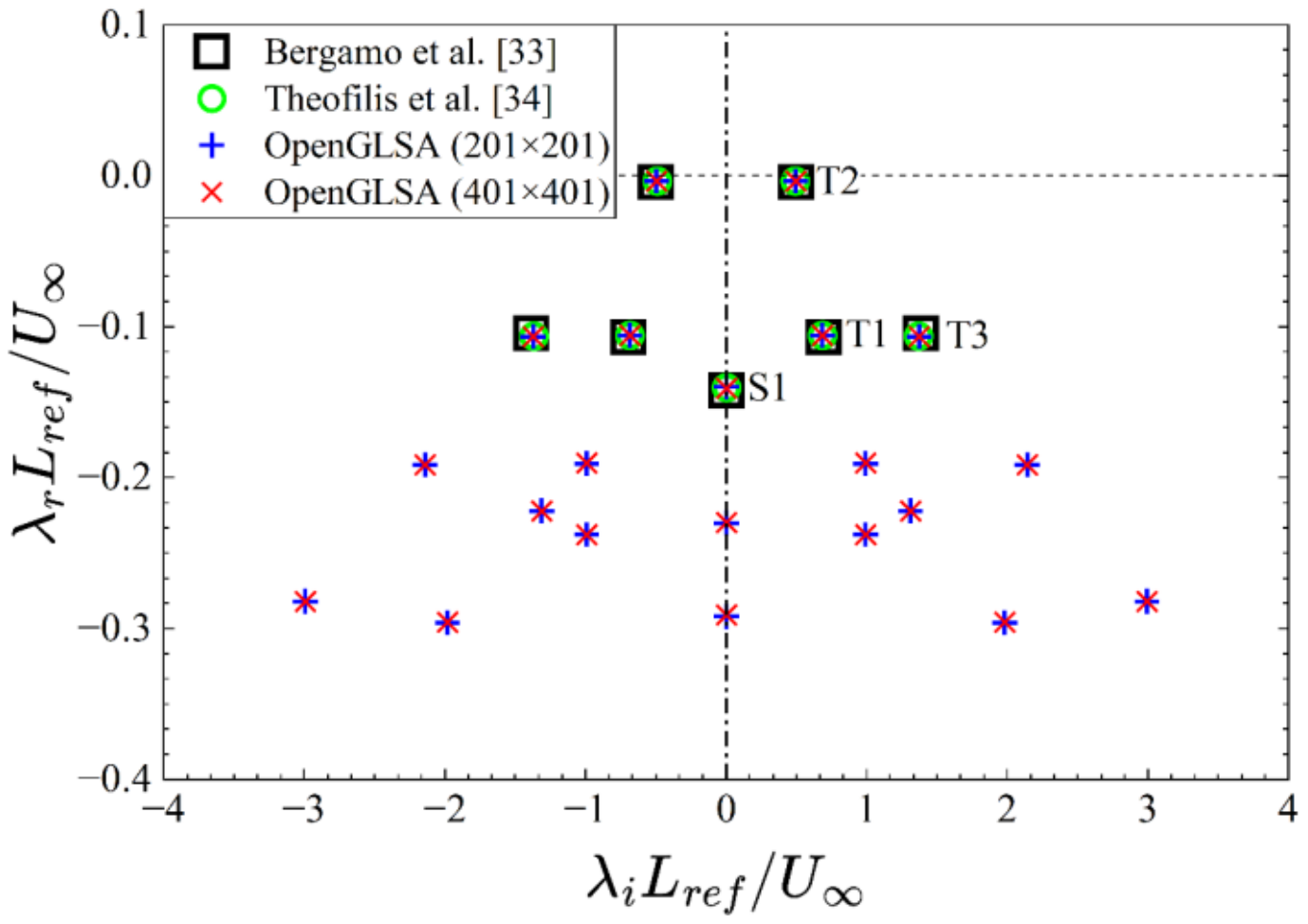

Figure 12 Eigenvalue spectrum of the square regularised lid-driven cavity flow with spanwise angular wavenumber $\beta = 7.35$. OpenGLSA results on (201×201) and (401×401) grids are compared with those of Bergamo et al. [33] and Theofilis et al. [34].

The converged eigenvalues are plotted in Figure 12. The results agree well with those reported in the references. The four least stable eigenvalues are listed in Table 1. The computed eigenvalues differ from the incompressible results of Theofilis et al. [34] by at most $O(10^{-2})$, and from the compressible results of Bergamo et al. [33] by at most $O(10^{-3})$. Refining the grid from (201×201) to (401×401) changes the eigenvalues by only $O(10^{-4})$, indicating weak sensitivity to the grid refinement considered here. Figure 13 shows the real parts of velocity perturbations of the least stable mode T2.

Table 1 Comparison with reference results. $Ma_\infty = 0.1$, $Re_\infty = 900$, and $\beta = 7.35$.

| | S1 | | T1 | | T2 | | T3 | |
|---|---|---|---|---|---|---|---|---|
| | $\pm\lambda_i$ | $\lambda_r$ | $\pm\lambda_i$ | $\lambda_r$ | $\pm\lambda_i$ | $\lambda_r$ | $\pm\lambda_i$ | $\lambda_r$ |
| Bergamo et al. [33] | 0.0000 | −0.1405 | 0.6855 | −0.1058 | 0.4925 | −0.0035 | 1.3691 | −0.1061 |
| Theofilis et al. [34] | 0.0000 | −0.1425 | 0.6928 | −0.1071 | 0.4981 | −0.0043 | 1.3846 | −0.1044 |
| OpenGLSA ($201^2$) | 0.0000 | −0.1400 | 0.6865 | −0.1056 | 0.4938 | −0.0036 | 1.3709 | −0.1063 |
| OpenGLSA ($401^2$) | 0.0000 | −0.1406 | 0.6868 | −0.1056 | 0.4936 | −0.0035 | 1.3715 | −0.1065 |

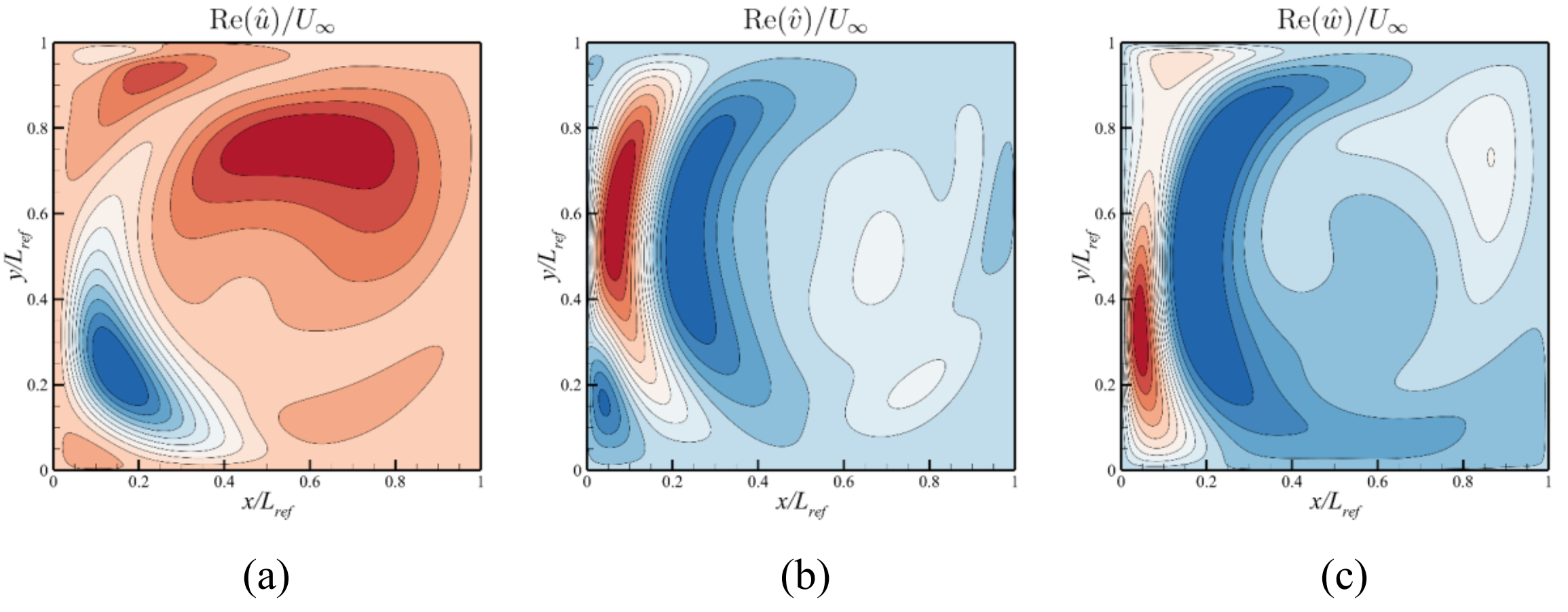

Figure 13 Real parts of velocity perturbations of the least stable eigenmode of the square regularised lid-driven cavity flow with spanwise angular wavenumber $\beta$ = 7.35. (a) x-velocity, (b) y-velocity and (c) z-velocity.

### *5.4. Cubic lid-driven cavity flow*

### *5.4.1. Base flow*

The cubic regularised lid-driven cavity flow is selected to test OpenGLSA's capability of performing the fully three-dimensional global stability analysis. The computational domain is a unit cube with one moving wall and five stationary walls. Following the settings used by Ohmichi and Yamada [36], the Mach number and Reynolds number based on the cube edge length $L_{\mathrm{ref}}$ = 1 are $Ma_\infty$ = 0.2 and $Re_\infty$ = 200, respectively. All walls are isothermal and the nondimensional wall temperature is 1. The pressure compatibility boundary is applied at all boundaries. The velocity distribution of the moving wall at $y$ = 1 is given as

$$u(x,1,z) = [1-(2x-1)^{16}]^2[1-(2z-1)^{16}]^2 . \tag{51}$$

The computational grid is a uniform Cartesian grid with 101 points in each direction. The same density correction procedure as in the square LDC case is applied, and the residual convergence history is plotted in Figure 14. The residual decreases below $10^{-11}$.

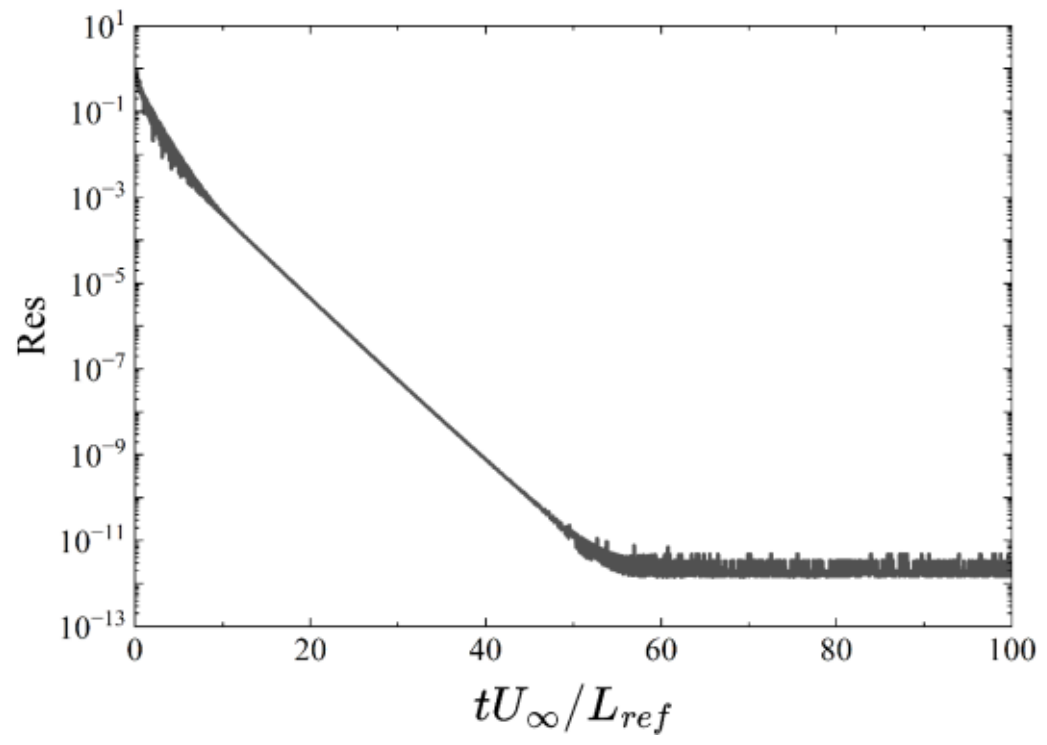

Figure 14 Convergence history of the residual of the cubic lid-driven cavity flow.

### *5.4.2. Global linear modes*

Global stability analysis is performed with a time step of $\Delta t$ = 5.0×10$^{-4}$ and a propagation interval of $\Delta T$ = 0.5. Twenty eigenvalues are requested and the dimension of the Krylov subspace is 30. The convergence tolerance is 10$^{-8}$. The calculation converges after 15 update iterations and 212 propagator evaluations, returning 21 converged Ritz values.

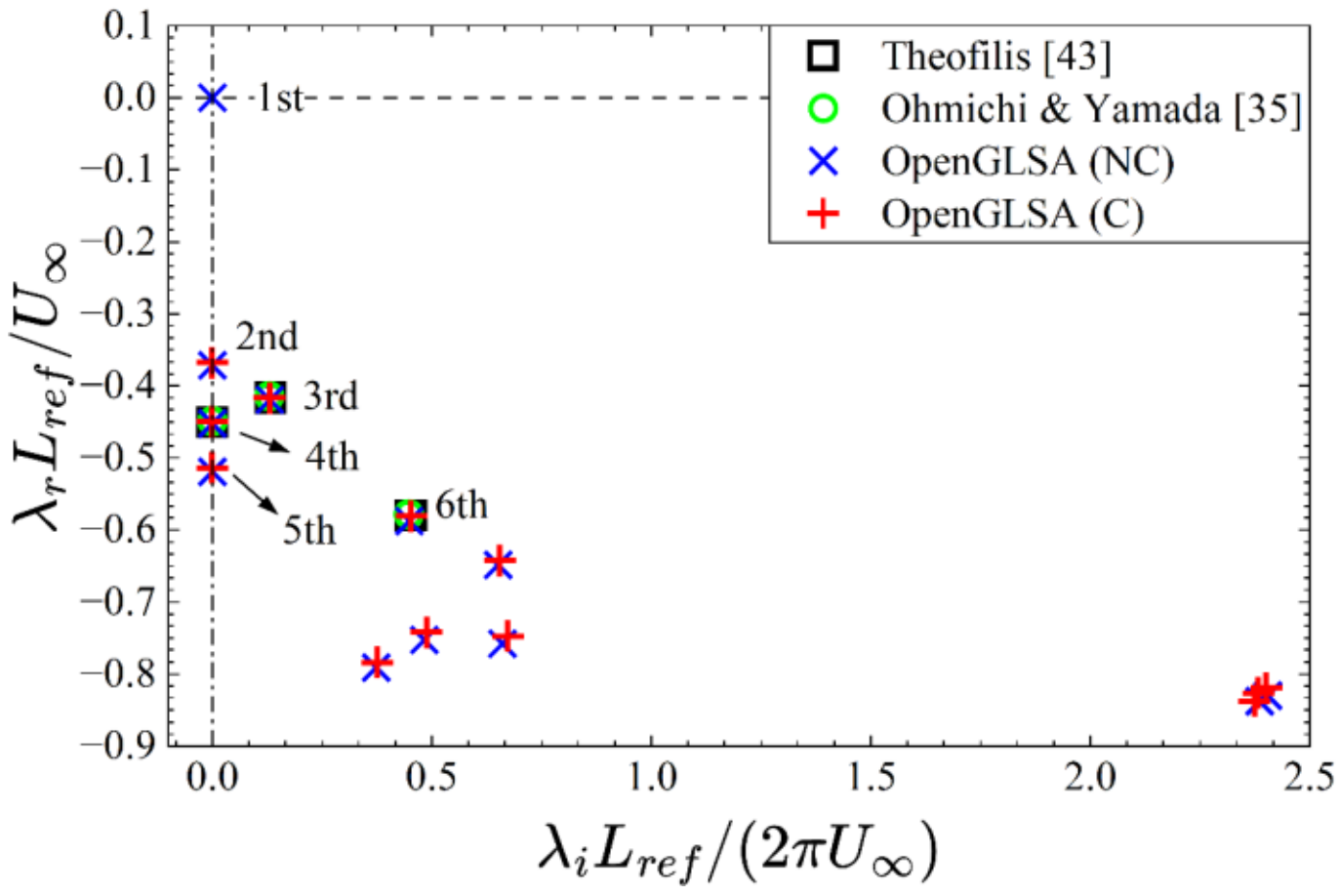

Figure 15 Eigenvalue spectrum of the cubic regularised lid-driven cavity flow. OpenGLSA results with (C) and without (NC) mass perturbation conservation correction are compared with those of Theofilis [35] and Ohmichi & Yamada [36].

Figure 15 compares the spectra obtained with and without the mass-perturbation correction with the results of Theofilis [35] and Ohmichi and Yamada [36]. Without the correction, an additional eigenvalue, labelled as the 1st mode in the figure, appears close to zero. This mode is associated with a non-zero volume integral of the density perturbation and therefore changes the total mass relative to the base flow. For perturbations that preserve the prescribed cavity mass, the admissible states satisfy

$$\int_\Omega \hat{\rho}\,\mathrm{d}V = 0 \tag{52}$$

and enforcing such a constraint removes the near-zero mass mode from the computed spectrum.

After the correction, two additional modes, labelled as the 2nd mode and 5th mode respectively, remain in the spectrum but are absent from the reference spectra. To characterise the computed modes, we use an equivalent pressure-entropy form of the compressible disturbance energy norm derived by Hanifi et al. [44]:

$$E_{Chu} = E_k + E_p + E_s, \tag{53}$$

where

$$\begin{cases} E_k = \dfrac{1}{2}\int_\Omega \bar{\rho}(|\hat{u}|^2 + |\hat{v}|^2 + |\hat{w}|^2) dV \\ E_p = \dfrac{1}{2}\int_\Omega \dfrac{|\hat{p}|^2}{\gamma \bar{p}} dV \\ E_s = \dfrac{1}{2}\int_\Omega \dfrac{\bar{\rho}\bar{T}}{(\gamma-1)Ma_\infty^2} |\hat{s}|^2 dV \end{cases} \tag{54}$$

are the kinetic, pressure and entropy contributions, respectively. The energy decomposition identifies them as entropy-dominated modes. As shown in Table 2, the second and fifth modes have entropy-energy fractions of 89.11% and 99.93%, respectively, whereas their kinetic-energy fractions are only 0.15% and 0.07%. The second mode also has a pressure-energy contribution of 10.74%. By comparison, kinetic energy accounts for 99.25%, 99.43% and 99.79% of the total energy of the third, fourth and sixth modes, respectively, which correspond to modes in the incompressible spectrum. The dominance of entropy energy in the two additional modes indicates that they are associated with the thermodynamic degrees of freedom introduced by the compressible formulation.

Table 2 Chu perturbation energy fractions for the five least stable modes of the cubic lid-driven cavity flow.

| | Mode No. | $E_k$ / $E_{\mathrm{chu}}$ (%) | $E_p$ / $E_{\mathrm{chu}}$ (%) | $E_s$ / $E_{\mathrm{chu}}$ (%) | Dominant |
|---|---|---|---|---|---|
| Incompressible modes | 3rd | 99.25 | 0.05 | 0.7 | Kinetic |
| | 4th | 99.43 | 0.12 | 0.45 | Kinetic |
| | 6th | 99.79 | 0.04 | 0.17 | Kinetic |
| Compressible modes | 2nd | 0.15 | 10.74 | 89.11 | Entropy |
| | 5th | 0.07 | 0.00 | 99.93 | Entropy |

Figure 16 shows iso-surfaces of the real parts of the momentum perturbation for the 3rd mode. These structures are consistent with those reported by Ohmichi and Yamada [36].

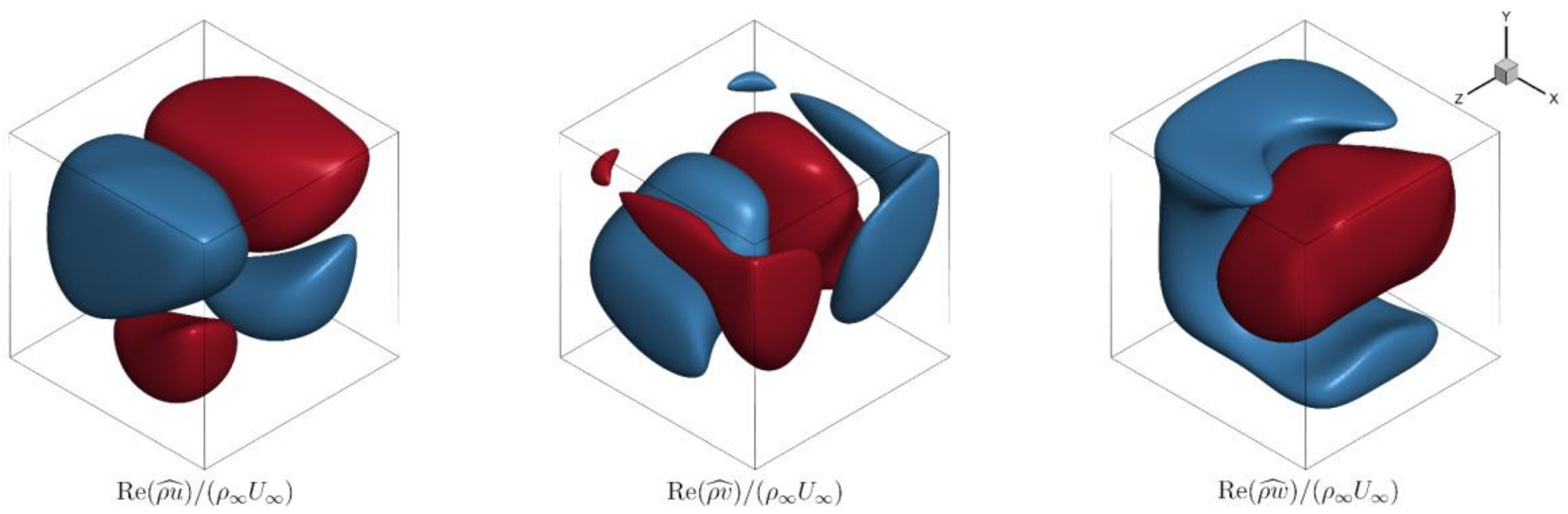

(a) (b) (c)

Figure 16 Kinetic-energy-dominated eigenmode of the cubic regularised lid-driven cavity flow (the 3rd mode, corresponding to the 1st direct mode in [36]). Iso-surfaces show the real parts of the (a) x-momentum perturbation, (b) y-momentum perturbation and (c) z-momentum perturbation.

This fully three-dimensional GSA requires approximately 11.52 GB of memory and a wall-clock time of around 13 hours.

### *5.5. Hypersonic ramp flow*

#### *5.5.1. Base flow*

A 15° compression ramp is considered to validate OpenGLSA for hypersonic flows. The Mach number is $Ma_\infty = 7.7$. A reference length of $L_{\mathrm{ref}} = 0.001$ m is used, giving a Reynolds number of $Re_\infty = 4.2\times10^3$. This corresponds to $Re_\infty = 4.2\times10^5$ based on the flat plate length of $L_{\mathrm{p}} = 0.1$ m in [6]. For comparison with the reference results, coordinates are presented using the plate length.

The computational grid is the same as that used in [6] with 1601 points along the streamwise direction and 321 points along the wall-normal direction. An isothermal no-slip wall boundary condition is prescribed with the wall temperature fixed at 293 K. Freestream conditions with $T_\infty = 125$ K are imposed at the inflow and upper boundaries. The outflow boundary uses an extrapolation condition. As the global instability is three-dimensional, the two-dimensional base flow is obtained by direct time integration without SFD.

The convective fluxes are reconstructed using characteristic-wise MR-WENO, and the viscous flux derivatives are evaluated using sixth-order central differences. The positivity-preserving limiter is applied during the base-flow computation.

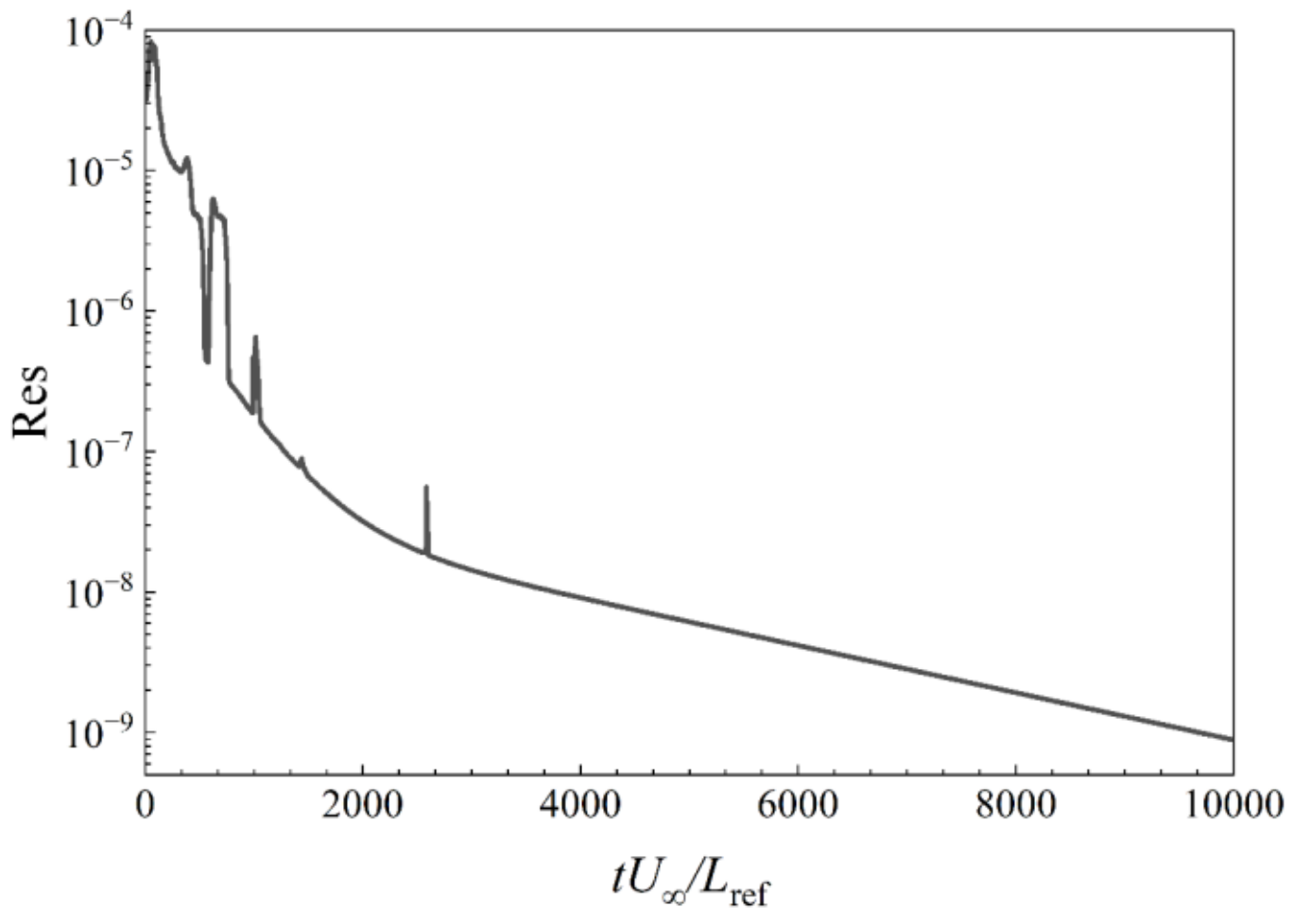

Figure 17 Convergence history of the residual for the hypersonic ramp base flow.

The convergence history of the residual is plotted in Figure 17. With the MR-WENO scheme, the residual decreases to the level around $10^{-9}$.

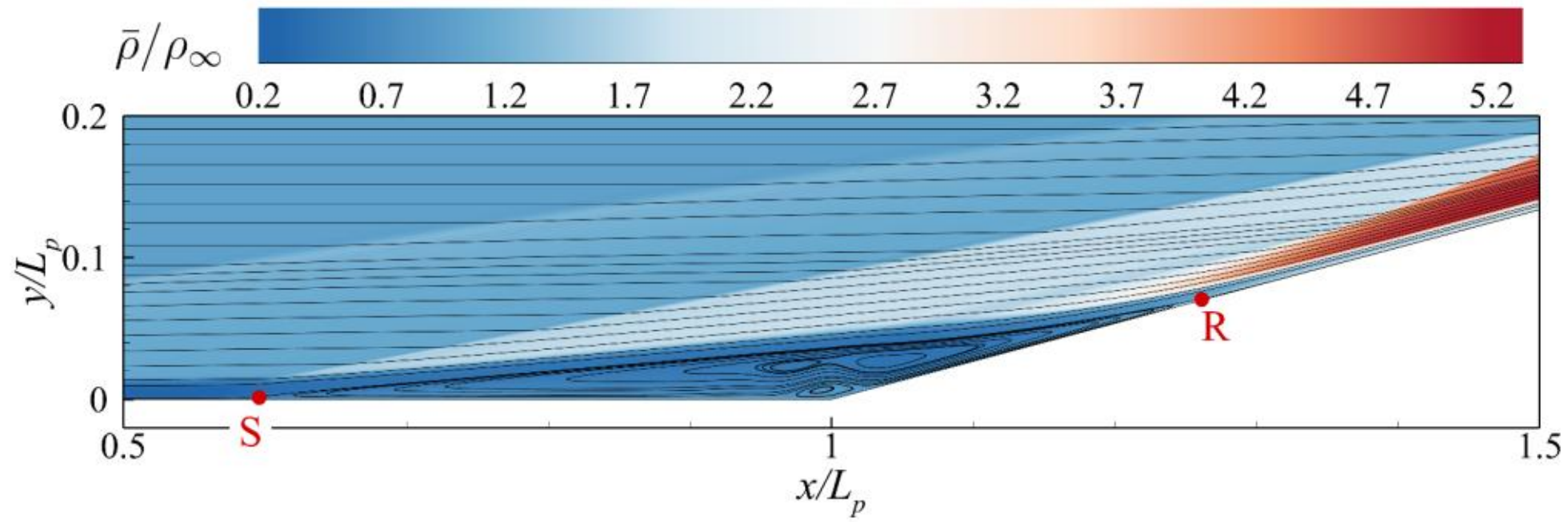

Figure 18 Density contour and streamlines of the hypersonic ramp flow. Red circles mark the separation point (S) and reattachment point (R).

The density contour and streamlines of the converged solution are plotted in Figure 18. The boundary layer separates upstream of the corner and reattaches on the ramp, forming a separation bubble with a streamwise extent of approximately $0.67L_p$. The separation and reattachment points are labelled *S* and *R*, respectively. A smaller secondary recirculation region is visible near the corner. Figure 19 shows the wall distributions of the pressure coefficient $C_p$ and skin-friction coefficient $C_f$, defined following [5] as

$$C_p = \frac{p_w}{0.5\rho_\infty U_\infty^2}, \qquad C_f = \frac{\tau_w}{0.5\rho_\infty U_\infty^2}, \tag{55}$$

where $\tau_w$ and $p_w$ are the wall shear stress and pressure, respectively. Both distributions agree closely with the reference results reported by Hao et al. [5]. The separation and reattachment locations identified from the sign changes of $C_f$ are also close to the reference values. Only small differences can be found near the downstream end of the ramp.

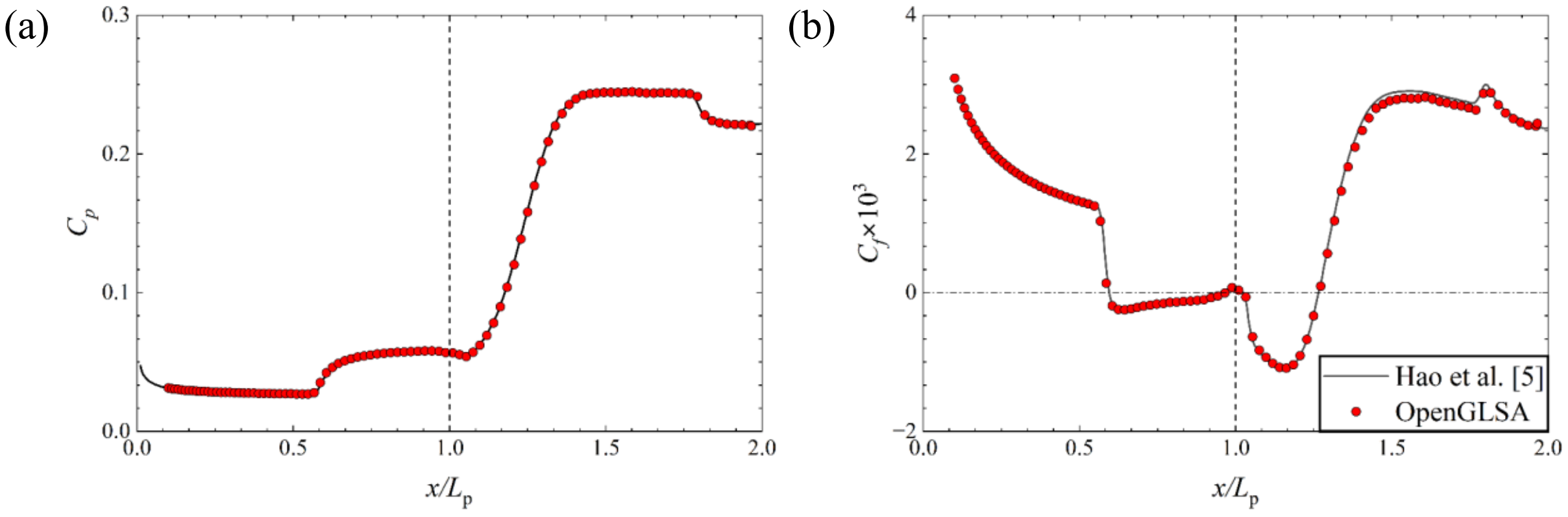

Figure 19  Wall distribution of (a) pressure coefficient $C_p$ and (b) skin-friction coefficient $C_f$ for the ramp flow. Comparison with Hao et al. [5].

### 5.5.2. Global linear modes

For global stability analysis, the propagation interval is set to $\Delta T = 2$, with a time-integration step of $\Delta t = 5.0\times10^{-4}$. The dimension of the Krylov subspace is set to 101, and the maximum number of restarts is 100. The relative convergence tolerance for the Ritz values is set to$10^{-8}$.

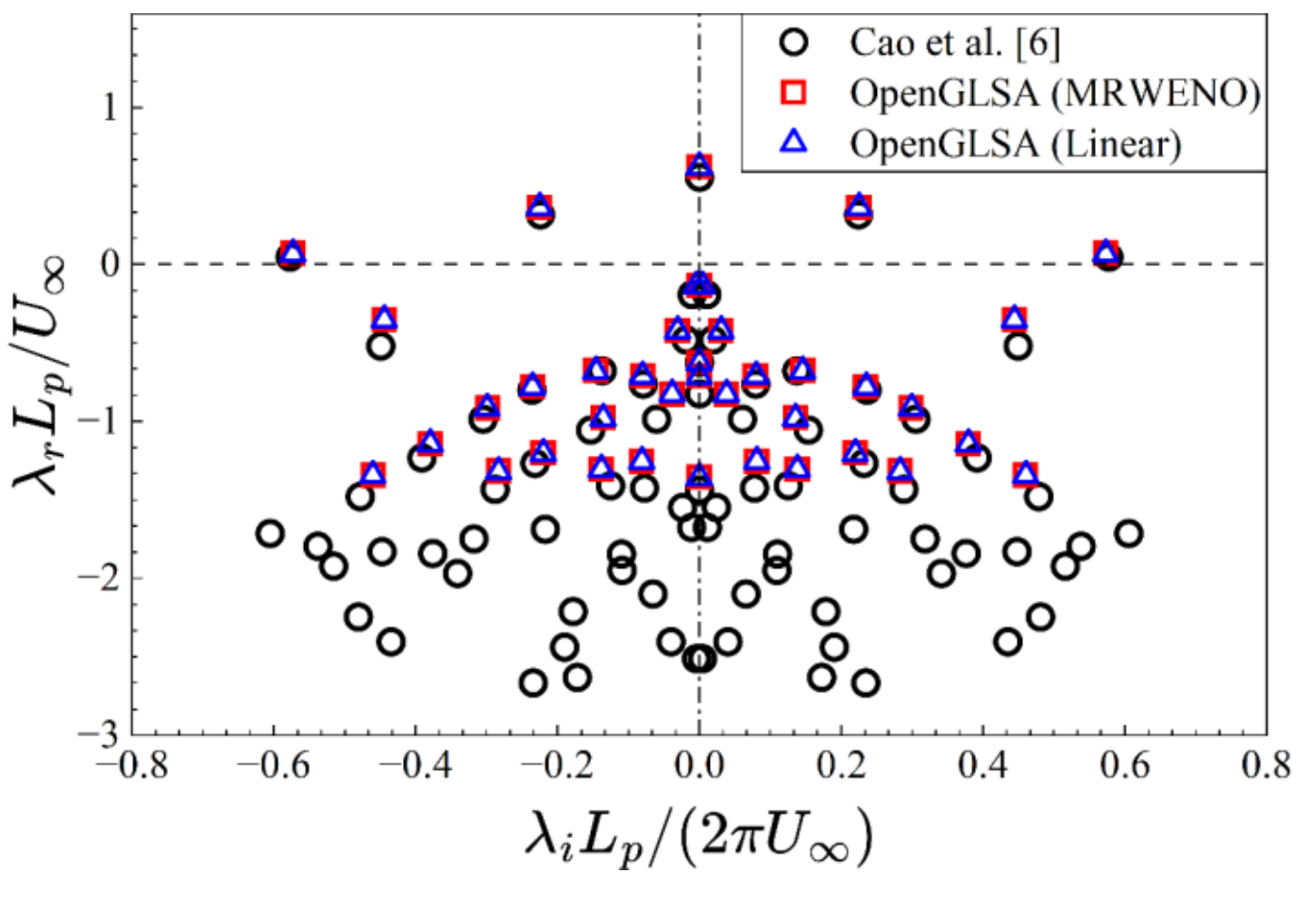

Figure 20  Comparison of eigenvalue spectra with Cao et al. [6] at spanwise angular wavenumber $\beta$ = 0.95 using the same nondimensionalisation as in [6], corresponding to the wavelength $L_z$ = 6.6 mm.

The spanwise angular wavenumber is set to $\beta$ = 0.95, corresponding to a physical wavelength of approximately 6.6 mm, at which the most unstable mode shows up. The computed spectra are compared with those reported by Cao et al. [6] in Figure 20. The three unstable branches, comprising one stationary mode and two pairs of oscillatory modes, show reasonable agreement with the reference results. The results obtained using MR-WENO and linear upwind reconstruction for the linearised equations are also compared and are nearly identical.

Figure 21 shows the real parts of the spanwise velocity perturbation for the three unstable modes identified in Figure 20. Mode 1 is stationary, with its strongest perturbations concentrated in the downstream part of the separation bubble near the corner. The oscillatory Modes 2 and 3 extend further upstream and exhibit alternating positive and negative regions along the streamwise direction, with more sign changes visible in Mode 3. These spatial features are consistent with the corresponding modes reported by Cao et al. [6].

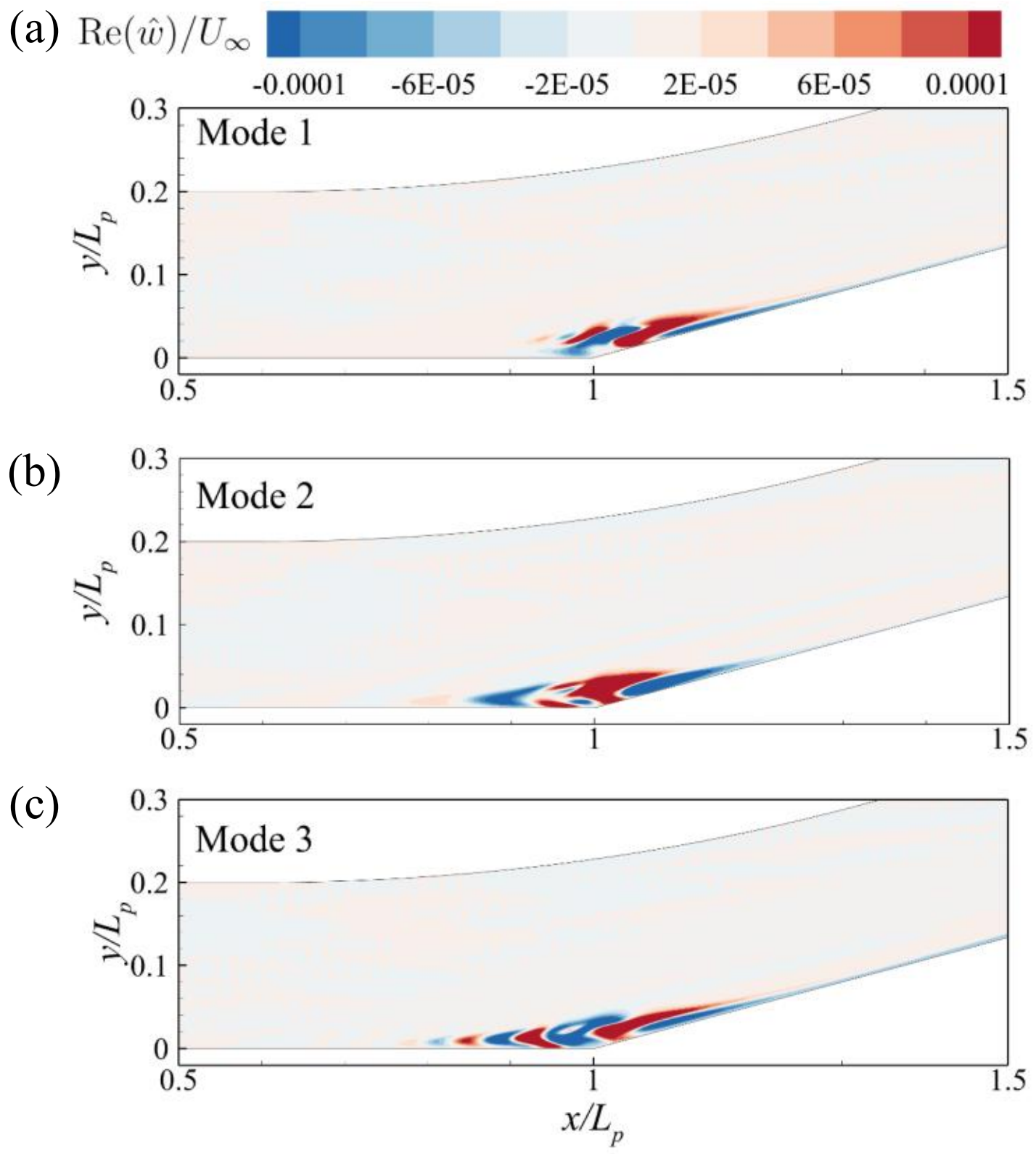

Figure 21 Real parts of the spanwise velocity perturbation $\hat{w}/U_\infty$ for the three most unstable eigenmodes of the hypersonic ramp flow: (a) stationary mode 1, (b) oscillatory mode 2 and (c) oscillatory mode 3.

To further examine the differences between the two reconstructions, density perturbation profiles of the most unstable stationary mode are compared at three stations as shown in Figure 22 (Modes are normalised to unit Chu energy). The profiles agree closely within the shear layer but diverge beyond the outer edges of the shear layer and shocks. At all three stations, the linear upwind profiles decay rapidly to near machine zero, consistent with the zero perturbations imposed by the freestream condition at the upper boundary. In contrast, the MR-WENO profiles retain small-amplitude oscillations, revealing differences in the spatial modes despite the nearly identical spectra.

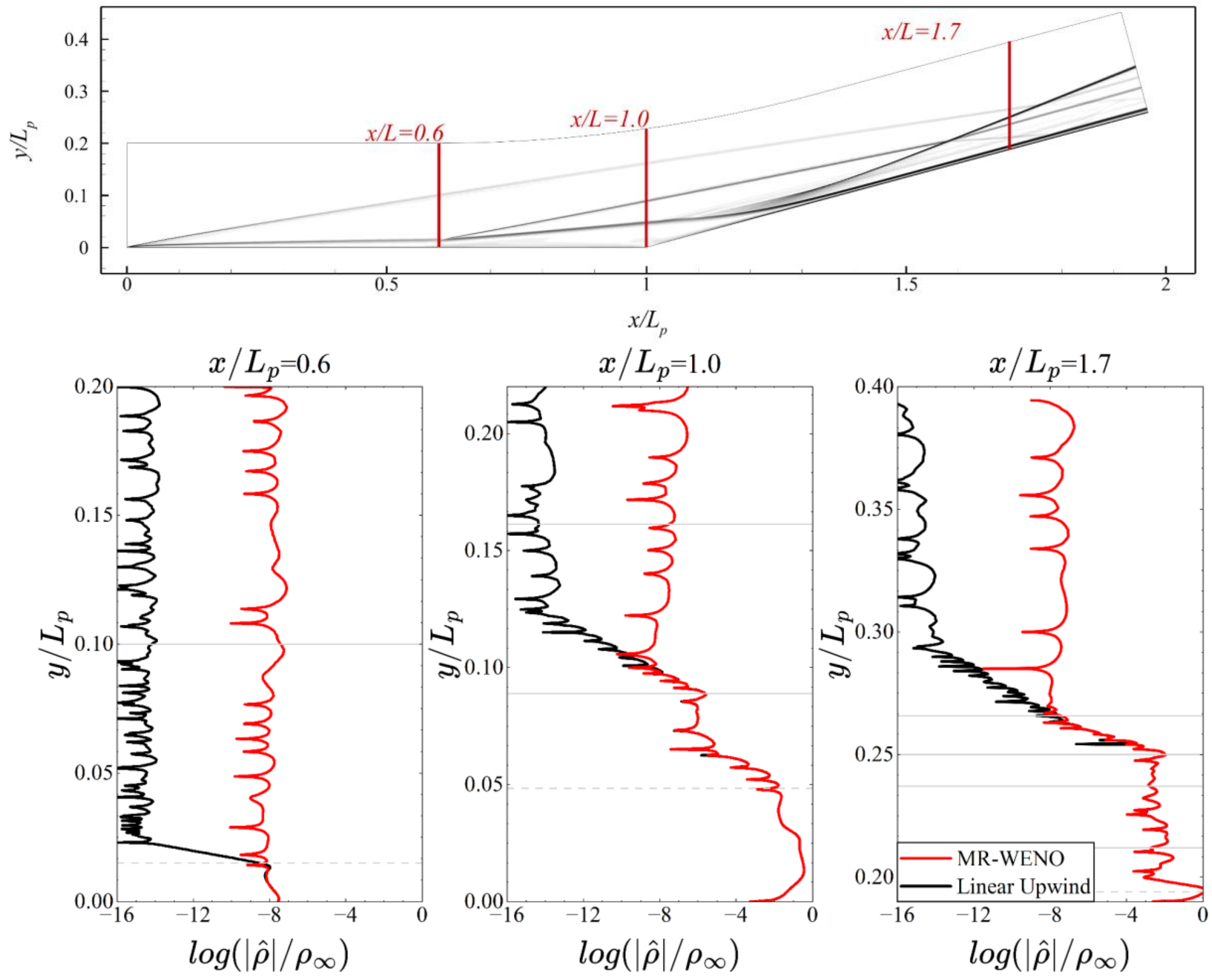

Figure 22 Numerical schlieren and density perturbation profiles of the most unstable eigenmode obtained using the MR-WENO and linear upwind schemes at $x / L_p = 0.6$, $x / L_p = 1.0$ and $x / L_p = 1.7$. Grey dashed lines denote the shear layer location; grey solid lines denote the shock location.

## 6. Conclusions and future work

We present the main algorithms and validation cases of the open-source code OpenGLSA. The code supports linear stability analysis around planar and axisymmetric two-dimensional base flows, with both two-dimensional perturbations and three-dimensional perturbations represented by Fourier mode expansions. It also supports fully three-dimensional linear stability analysis in general curvilinear grids. The validation cases cover low-Mach-number and hypersonic external flows, as well as square and cubic lid-driven cavity flows. The computed base flows and stability results agree well with published reference data for the cylinder, sphere and cavity cases, and reasonably for the hypersonic ramp case. By avoiding explicit assembly and factorisation of the Jacobian matrix, the matrix-free time-stepping approach used reduces memory requirements and enhancing both the efficiency and tractability of memory-intensive compressible flow analyses, although the cylinder comparison shows longer computation times than matrix-storage algorithms in BROADCAST under the tested case.

The current implementation uses single-block structured grids. Future work will focus on multiblock support. Turbulence models are currently being implemented. An adjoint perturbation solver is also planned to support sensitivity and resolvent analyses.

## CRediT authorship contribution statement

**Rui-Qing Yang**: Conceptualization, Methodology, Software, Validation, Formal analysis, Investigation, Data curation, Visualization, Writing – original draft.

**Bao-Feng Ma**: Conceptualization, Methodology, Supervision, Investigation, Writing – review & editing.

## Declaration of generative AI and AI-assisted technologies in the manuscript preparation process

During the preparation of this work the authors used DeepSeek in order to improve the language and readability of the manuscript. After using this tool, the authors reviewed and edited the content as needed and take full responsibility for the content of the published article.

## Declaration of competing interest

The authors declare that they have no known competing financial interests or personal relationships that could have appeared to influence the work reported in this paper.

## Data availability

Data will be made available on request.

## Acknowledgements

The project was supported by the National Natural Science Foundation of China under Grant Nos.12072015 and 12572259. The authors thank the developers of the open-source code OpenCFD-SC. The code organisation, MPI parallelisation and parts of the nonlinear module of OpenGLSA were adapted from OpenCFD-SC.